\documentclass[5p,twocolumn]{elsarticle}
\usepackage{amsmath,graphicx}
\usepackage{subfigure}
\usepackage{lineno}
\usepackage{multirow}
\usepackage{stfloats}
\usepackage{amsfonts}
\usepackage{amssymb}
\usepackage{booktabs} 
\usepackage{hyperref}
\usepackage{color}
\usepackage{caption}

\newtheorem{proposition}{\hspace{0em}Proposition}

\newtheorem{remark}{\hspace{0em}Remark}

\modulolinenumbers[5]
\allowdisplaybreaks[4]

\begin{document}

\captionsetup[figure]{labelfont={bf},labelformat={default},labelsep=period,name={Fig.}}

\begin{frontmatter}

\title{Multi-sensor joint target detection, tracking and classification via Bernoulli filter}

\author [a]{Gaiyou Li}
\ead{lgyuestc@163.com}

\author [a]{Ping Wei}
\ead{pwei@uestc.edu.cn}

\author [b]{Giorgio Battistelli}
\ead{giorgio.battistelli@unifi.it}

\author [b]{Luigi Chisci}
\ead{luigi.chisci@unifi.it}

\author [a]{Lin Gao}
\ead{lingao\_1014@126.com}

\address[a]{School of Information and Communication Engineering,  University of Electronic Science and Technology of China,  Chengdu, Sichuan 611731, China}

\address[b]{Dipartimento di Ingegneria dell'Informazione, Universit\`a degli Studi di Firenze}



\begin{abstract}
	This paper focuses on \textit{joint detection, tracking and classification} (JDTC) of a target via multi-sensor fusion. 
	The target can be present or not, can belong to different classes, and depending on its class can behave according to different kinematic modes.
	Accordingly, it is modeled as a suitably extended Bernoulli \textit{random finite set} (RFS) uniquely characterized by existence, 
	classification, class-conditioned mode  
	and class\&mode-conditioned state probability distributions. 
	By designing suitable centralized and distributed rules for fusing information on target existence, class, mode and state from different sensor nodes,
	novel \textit{centralized} and \textit{distributed} JDTC \textit{Bernoulli filters} (C-JDTC-BF and D-JDTC-BF),
	are proposed.
	The performance of the proposed JDTC-BF approach is evaluated by means of simulation experiments.
\end{abstract}

\begin{keyword}
Distributed target tracking, 
target classification, 
random finite set,
Bernoulli filter,
information fusion
\end{keyword}

\end{frontmatter}



\section{Introduction}

Target detection and tracking are crucial tasks in surveillance (e.g., radar \cite{khan1994target},
sonar \cite{yocom2011bayesian}, 
autonomous driving \cite{petrovskaya2009model}, 
mobile robotics \cite{robin2016multi}) systems.
Though a possible approach is to separately deal with detection and tracking as two sequential phases \cite{clarke1996detection}, 
i.e. by first performing
target initialization followed by track maintainance,
it has been recognized that joint processing of both tasks can substantially enhance the overall performance \cite{vo2012multi}. 
In certain circumstances, 
it is also desired to perform target classification for high-level applications aiming at situation assessment \cite{bar2005tracking}. 
Moreover, knowledge of the target class provides valuable information on the possible kinematic behaviors of the target \cite{gordon2002efficient} (e.g., a fighter aircraft can perform sharper maneuvers than a cargo aircraft) which, in turn, can be profitably exploited
to improve tracking performance
\cite{challa2001joint}. 
The recent development of \textit{random finite set} (RFS) methods has produced
several interesting contributions to
\textit{joint detection, tracking and classification} (JDTC) of both a single and multiple targets \cite{yang2014joint,yang2015joint,li2012joint,yang2013joint,li2016multi,gao2016extensions,wei2012joint} but all 
based on a single-sensor system.

In many practical scenarios,
however, multi-sensor surveillance systems
entail significant advantages in terms of enhanced tracking accuracy,
target observability \cite{ilic2013adaptive}, reliability as well as
expanded coverage \cite{olfati2008distributed},
thus motivating the interest of the present work
for multi-sensor JDTC.
In this respect, the attention will be devoted to both a
centralized configuration wherein a fusion center gathers
measurements from all sensors,
or a distributed one
wherein each sensor updates a local posterior with its own measurements 
and then fuses it with the posteriors of the neighbors.
It is well-recognized that
centralized fusion can provide better performance while, on the other hand, distributed fusion is preferable 
in terms of 
scalability and fault tolerance \cite{ristic2013target}.

The goal of this paper is to address single-target multi-sensor JDTC,
exploiting both centralized and distributed configurations.
In order to
account for target appearance/disappearance and the presence of clutter,
the target is modeled as a  Bernoulli RFS \cite{guldogan2014consensus}.
Then, for classification purposes, different target classes are considered, each being characterized by a different set of possible kinematic modes.
Overall, target's information, to be recursively propagated in time, consists of \textit{existence probability} (EP), \textit{class probability mass function} (CPMF),
class-conditioned (kinematic) 
\textit{mode PMFs} (MPMFs) and class\&mode-conditioned \textit{state probability density functions} (SPDFs).
It is worth to point out that joint estimation of class and mode positively affects both classification and tracking.
In fact, kinematic mode information
can be exploited for target classification and,
inversely,
class information can help to define the possible modes of a target,
thus improving tracking performance \cite{yang2014joint}.

For the centralized configuration,
the optimal posterior based on measurements from all sensors
is obtained at each iteration
following the Bayesian approach.
For the distributed configuration, on the other hand,
the local posterior of each sensor is first obtained 
with the existing Bernoulli-JDTC method \cite{yang2015joint},
and then \textit{generalized covariance intersection} (GCI)  
\cite{uney2013distributed,battistelli2015distributed,wang2016distributed,clark2010robust} is exploited to fuse the local posteriors
so as to achieve EP, CPMF, MPMFs and SPDFs of the global posterior.
Furthermore,
\textit{Gaussian mixture} (GM) implementation of the proposed method is provided.

Finally,
simulation experiments are carried out
to comparatively evaluate performance of the proposed algorithms.
Summarizing, the main contributions of this paper are listed as follows:
\begin{itemize}
	\item the problem of multi-sensor JDTC is addressed under both centralized and distributed settings;
	\item the GM implementation of both algorithms, 
	as opposed to the particle filter implementation adopted in \cite{yang2015joint} for the single-sensor JDTC Bernoulli filter, 
	is discussed.
\end{itemize}

Notice that,
to the best of our knowledge,
this is the first paper addressing distributed JDTC. 
The rest of this paper is organized as follows. 
Section 2 formulates the multi-sensor JDTC problem considered in this paper.
Sections 3 and 4  present novel centralized and, respectively, distributed multi-sensor JDTC Bernoulli filters.
Section 5 deals with the Gaussian-mixture implementation of the proposed filters.
Section 6 evaluates the performance of the proposed filters by simulation experiments.
Finally,
Section 7 ends the paper with concluding remarks.

\section{Problem formulation}

This section formulates the JDTC problem of interest, relative to a single target in a clutterly environment surveiled by multiple sensors.
We will first model target dynamics and then multi-sensor measurement generation.

\subsection{Target dynamics}
To represent the target, let us introduce the following three items:
\textit{kinematic state} $x \in \mathbb{X}$, $\mathbb{X}$ being an Euclidean state space;
\textit{class} $c \in \mathcal{C}$, $\mathcal{C}= \lbrace{c_1, \cdots, c_{|{\cal C}|}}\rbrace$ being a discrete class set;
\textit{(kinematic) mode} $m \in \mathcal{M}_c$, $\mathcal{M}_c = \{m_{c,1}, \cdots, m_{c,{|\mathcal{M}_c|}}\}$ being a class-dependent discrete mode set, where $| \cdot |$ denotes \textit{cardinality} .
To summarize target information it is, therefore, convenient to define the augmented state vector
$\mathbf{x}=[x^{\top}, c,  m]^{\top} \in {\mathbb{X}} \times \mathcal{C} \times \mathcal{M}_c   $.

Since the target can either exist or not, it is naturally modeled as a Bernoulli RFS which can be either empty or a singleton, 
in the augmented state space,
 with some \textit{existence probability} (EP)
$r \in [0,1] $.
Accordingly, the target set density is defined as
\begin{equation}
f (\mathcal{X}) = \left\{ \begin{array}{ll} 1-r, & \mbox{if } \mathcal{X} = \emptyset \\
r s ( \mathbf{x} ), & \mbox{if } \mathcal{X} = \{ \mathbf{x} \} \\
0, & \mbox{if } |\mathcal{X} | > 1
\end{array}
\right.
\label{BD}
\end{equation}
where $s(\mathbf{x}) = s (x,c,m)$ is the augmented \textit{state PDF}.
By the \textit{chain rule}, such a PDF can be factored as
\begin{equation}
s(\mathbf{x}) = s(x,c,m) = s(x|c,m) \, \beta(m|c) \, \gamma(c)
\label{JSPDF}
\end{equation}
where:
$\gamma(c)$ is the \textit{class PMF} (CPMF); 
$\beta(m|c)$ is the class-conditioned \textit{mode PMF} (MPMF);
$s(x|c,m)$ is the class\&mode-conditioned \textit{state PDF} (SPDF).
Hence, the target is completely characterized by EP $r$, CPMF $\gamma(\cdot)$, MPMFs $\{ \beta(\cdot | c) \}_{c \in \mathcal{C}}$
and SPDFs $\{ s(\cdot|c,m) \}_{c \in \mathcal{C}, m \in \mathcal{M}_c}$.
For the sake of simplicity, hereafter the Bernoulli set density (\ref{BD})-(\ref{JSPDF}) will be referred to with the shorthand notation $f = \left\{ r, \gamma, \beta, s \right\}$.

The target dynamics has to account for appearance (birth), disapperance (death) and motion.
In this respect, it can be completely characterized by the transition density $\Phi_{k|k-1} \left( \mathcal{X}_+ | \mathcal{X} \right)$ which expresses, in probabilistic terms,
the transition of the target set from $\mathcal{X}$ at time $k-1$ to $\mathcal{X}_+$ at time $k$.
Specifically,
\begin{align}
 \Phi_{k|k-1}(\mathcal{X}_+|\emptyset)=
 \begin{cases}
 	1-p_B, & \mbox{if } \mathcal{X}_+ = \emptyset \\
 	p_B\cdot b(\mathbf{x}_+), & \mbox{if } \mathcal{X}_+ = \lbrace \mathbf{x}_+ \rbrace
 \end{cases} \label{eq3} \\
 \Phi_{k|k-1}(\mathcal{X}_+| \{ {\mathbf{x}} \})=
 \begin{cases}
 1-p_S, & \mbox{if } \mathcal{X}_+ = \emptyset \\
 p_S\cdot \phi_{k|k-1}(\mathbf{x}_+|\mathbf{x}), & \mbox{if } \mathcal{X}_+ = \lbrace \mathbf{x}_+ \rbrace
 \end{cases} \label{eq4} 
\end{align}
where: 
$p_B$ and $p_S$ are the probabilities of appearance of a newborn target and, respectively, of survival of an existing target;
$b(\mathbf{x}_+)=\gamma_B(c_+) \, \beta_B(m_+|c_+) \, s_B(x_+|c_+,m_+)$ is the a priori state PDF of the potential new target;
$\phi_{k|k-1}(\mathbf{x}_+|\mathbf{x})$ is the transition PDF, in the augmented state space, 
of the existing target. 
For ease of presentation, the survival probability $p_S$ is supposed to be independent from the (augmented) state, 
but all the ensuing developments can be readily generalized to the case of a state-dependent survival probability.
For the target class, mode and state evolution, the following reasonable assumptions are made:
\begin{itemize}
\item the target class remains constant over time, i.e. $c_k = c_{k-1}$;
\item the mode transition is governed by a class-dependent homogeneous Markov chain with transition probabilities
\begin{equation}
Prob \left( m_k = m_+ | m_{k-1} = m, c \right) = \pi_c(m_+|m);
\label{tp}
\end{equation}
\item the target motion is modeled by the mode-dependent 
state transition density
\begin{equation}
Prob \left( x_k = x_+ | x_{k-1} = x, m_{k} = m \right)  = \varphi(x_+ | x, m ) \, .
\label{motion}
\end{equation}

\end{itemize}
Taking into account the above assumptions, the transition PDF turns out to be given by
\begin{equation}
\begin{array}{rcl}
\Phi_{k|k-1}(\mathbf{x}_+ | \mathbf{x})  & = &
\Phi_{k|k-1}(x_+, c_+, m_+ | x, c, m ) \vspace{1mm} \\
 & = & \delta_{c_+,c} \, \pi_c(m_+|m) \, 
  \varphi(x_+ | x, m_+ )
 \end{array} 
 \label{eq7}
\end{equation}
where $\delta_{c_+,c}$ is the \textit{Kronecker delta} equal to $1$ if $c_+ = c$ and to zero otherwise.

\subsection{Multi-sensor measurement model}
The area of interest is monitored by a set of sensors $\mathcal{N} = \{ 1, \dots, |\mathcal{N} | \}$.
At sampling time $k$, sensor node $i \in \mathcal{N}$ provides
the measurement RFS  
\begin{align}
\mathcal{Z}_k^i= \mathcal{T}^i (\mathcal{X}_k) \, \cup \, \mathcal{K}^i_k, \label{Node_Meas}
\end{align}
which is the union of the target-originated RFS $\mathcal{T}^i (\mathcal{X}_k)$ and the clutter set $\mathcal{K}^i_k$.
The target-originated RFS takes the form
\begin{equation}
\begin{array}{rcl}
\mathcal{T}^i (\mathcal{X}_k) & = & 
\begin{cases}
\emptyset, & \mbox{if } \mathcal{X}_k = \emptyset \\
\emptyset, & \mbox{if } \mathcal{X}_k = \lbrace \mathbf{x}_k \rbrace \mbox{ with prob. } 1-p_D^i(x_k,c)\\
z^i_k, & \mbox{if } \mathcal{X}_k = \lbrace \mathbf{x}_k \rbrace \mbox{ with prob. } p_D^i(x_k,c) 
\end{cases} \vspace{2mm} \\
z^i_k & \sim & \ell^i(z_k^i|x_k) 
\end{array}
\label{target-meas}
\end{equation}
where $ \ell^i(z|x) $ is the likelihood function associated to the $i^{th}$ sensor.
The clutter set $\mathcal{K}^i_k$ is modeled as Poisson RFS \cite{mahler2007statistical} with 
\textit{probability hypothesis density} (PHD) $\kappa(z)$ defined over the measurement space. The measurements of different sensors
are supposed to be mutually conditionally independent.
Notice that, in the considered multi-sensor measurement model, the target class $c \in {\cal C}$ only affects detection probabilities $p_D^i(x,c)$ while the target mode $m \in {\cal M}_c$ is irrelevant.

\section{Centralized JDTC-BF algorithm}

This section focuses on the centralized configuration wherein all sensor nodes $i \in {\cal N}$ convey their measurement sets $\mathcal{Z}_k^i$
to a fusion center that,
in principle, 
should be able to perform optimal multi-sensor fusion, i.e. to provide the Bernoulli set density
$f (\mathcal{X}_k |\bigcup\nolimits_{i \in \mathcal{N}} {\mathcal{Z}_{1:k}^i})$
where ${\mathcal{Z}_{1:k}^i}$ denotes the sequence of measurements collected by sensor $i$ from time $1$ to time $k$.
Hereafter, it will be shown how to extend the joint detection and tracking Bernoulli filtering approach of 
\cite{vo2012multi,ristic2013target,mahler2007statistical}
to the JDTC setting of this paper.
Specifically, assuming that at time $k-1$ the augmented Bernoulli density $f_{k-1} = \{ r_{k-1}, \gamma_{k-1}, \beta_{k-1}, s_{k-1} \}$ is given and following a Bayesian approach, the aim is to first perform prediction to obtain $f_{k|k-1} = \{ r_{k|k-1}, \gamma_{k|k-1}, \beta_{k|k-1}, s_{k|k-1} \}$ by exploiting the target dynamics (\ref{eq3})-(\ref{eq7}), then followed by multi-sensor update to get $f_{k} = \{ r_{k}, \gamma_{k}, \beta_{k}, s_{k} \}$ by exploiting the measurement model (\ref{Node_Meas}).

\subsection {JDTC-BF prediction}

Prediction of a standard Bernoulli RFS density $f_{k-1} = \{ r_{k-1}, s_{k-1} \}$ into $f_{k|k-1} = \{ r_{k|k-1}, s_{k|k-1} \}$ can be found in \cite[Eqs. (10)-(12)]{ristic2013target}.
The following result provides the extension of such prediction to a Bernoulli RFS density defined over the augmented class-mode-state space.

\begin{proposition} 
	Given
	the Bernoulli RFS density $f_{k-1}=\lbrace r_{k-1},\gamma_{k-1},\beta_{k-1},s_{k-1} \rbrace $,
	the predicted
	density $f_{k|k-1}=\lbrace r_{k|k-1},\gamma_{k|k-1},\beta_{k|k-1} ,s_{k|k-1}\rbrace$ is obtained as follows
		\begin{align}
		&r_{k|k-1} =p_B(1-r_{k-1}) +p_S \, r_{k-1}  , \label{pred1} \\
		&\gamma_{k|k-1}(c)=  \frac{p_B(1-r_{k-1})}{r_{k|k-1}} \gamma_B(c) 
		+\frac{ p_S \,  r_{k-1}}{r_{k|k-1}} \gamma_{k-1}(c)  ,\label{pred2} \\
		&\beta_{k|k-1}(m|c)=\frac{p_B(1-r_{k-1})}{r_{k|k-1}\gamma_{k|k-1}(c)} \gamma_B(c)\beta_B(m|c) \nonumber \\ 
		&+\frac{p_S \, r_{k-1} }{r_{k|k-1}\gamma_{k|k-1}(c)} \gamma_{k-1}(c) \sum\limits_{m' \in \mathcal{M}_c}\pi_c(m|m')\beta_{k-1}(m'|c) , \label{pred3} \\
		&s_{k|k-1}(x|c,m) \nonumber\\
		&= \frac{p_B(1-r_{k-1}) \, \gamma_B(c) }{r_{k|k-1}\gamma_{k|k-1}(c)\beta_{k|k-1}(m|c)} \, \beta_B(m|c) \, s_B(x|c,m) \nonumber\\
		&+\frac{p_S \, r_{k-1} \gamma_{k-1}(c)  }{r_{k|k-1}\gamma_{k|k-1}(c)\beta_{k|k-1}(m|c)} \sum\limits_{m' \in \mathcal{M}_c}\pi_c(m|m') \nonumber \\
		& \quad \times  \beta_{k-1}(m'|c) \int  \varphi(x|x',m)  \, s_{k-1}(x'|c,m')dx'.\label{pred4} 	
	\end{align}
	
\end{proposition}
\textit{Proof:} see Appendix $A$. \qed

\subsection{JDTC-BF centralized multi-sensor update}
The following result concerns the centralized multi-sensor update of the augmented target Bernoulli density

\begin{proposition}
	Given
	the predicted Bernoulli RFS density $f_{k|k-1}=\lbrace r_{k|k-1},\gamma_{k|k-1},\beta_{k|k-1},s_{k|k-1} \rbrace $ and the measurement RFSs $\left\{ {\cal Z}_k^i \right\}_{i \in {\cal N}}$,
	the updated 
	density $f_{k}=\lbrace r_{k},\gamma_{k},\beta_{k} ,s_{k}\rbrace$ is obtained as follows
		\begin{align}
		r_k&= \frac{r_{k|k-1}  \sum\limits_{c \in {\cal C}} {\gamma}_{k|k-1}(c) \,  \ell(c) }{1-r_{k|k-1}+r_{k|k-1}\sum\limits_{c \in {\cal C}} {\gamma}_{k|k-1}(c) \,  \ell(c)}, \label{MS_EP_Context} \\
		\gamma_{k}(c)&= \frac{{\gamma}_{k|k-1}(c) \, \, \ell(c) }{\sum\limits_{c \in {\cal C}} {\gamma}_{k|k-1}(c) \, \ell(c)}, \\
		{\beta}_{k}(m|c)&= \frac{{\beta}_{k|k-1}(m|c) \, \ell(m|c)}{\sum\limits_{m \in \mathcal{M}_c} {\beta}_{k|k-1}(m|c) \, \ell(m|c)}, \label{MS_MPMF_CPMF_Context} \\
		s_k(x|c,m) 
		&  = \frac{s_{k|k-1}(x|c,m) \, \ell(x|c,m)}{\int s_{k|k-1}(x'|c,m) \, \ell(x'|c,m) \, dx'} \, , 
		\label{MS_SPDF_GM_Context}
	\end{align}
	where
		\begin{align}
	    \ell(c) & =  \sum\limits_{m \in \mathcal{M}_c} {\beta}_{k|k-1}(m|c) \, \ell(m|c) \, , \label{TC_LLF_GM_Context} \\
	    \ell(m|c) &= \int s_{k|k-1}(x|c,m) \, \ell(x|c,m) \, dx , \label{TM_LLF_GM_Context}  \\
	    \ell(x|c,m)&= \prod\limits_{i \in \mathcal{N}}  \Bigg [ 1-p^i_D(c) \nonumber \\
	    &+ p^i_D(c) \sum\limits_{z \in \mathcal{Z}^i_k} \dfrac{ \ell^i(z|x)}{\kappa(z)} \Bigg ] \, .   \label{TS_LLF_GM_Context} 
	\end{align}

\end{proposition}
\textit{Proof:} see Appendix $B$. \qed

\begin{table}[htbp] 
	\caption{\label{tab:test}} 
	\begin{tabular}{lcl} 
		\toprule 
		\textbf{Algorithm} C-JDTC-BF  (Time $k$) \\ 
		\midrule 
		$\mathbf{Input}$: $\mathcal{Z}_k=\lbrace \mathcal{Z}^i_k\rbrace_{i \in \mathcal{N}},f_{k-1}=\lbrace r_{k-1},\gamma_{k-1},\beta_{k-1},s_{k-1}\rbrace$  \\

		$\mathbf{Prediction}$: \\
		Predicted EP $r_{k|k-1}$, CPMF $\gamma_{k|k-1}(c)$, MPMFs \\
		 $\beta_{k|k-1}(m|c)$ and 
		SPDFs $s_{k|k-1}(x|m,c)$  
		are computed \\ by (\ref{pred1})-(\ref{pred4}).\\
		$\mathbf{Multi-sensor \quad update}$: \\
		Updated EP $r_{k}$, CPMF $\gamma_{k}(c)$, MPMFs $\beta_{k}(m|c)$ \\
		 and SPDFs $s_{k}(x|c,m)$  
		are computed by 
		(\ref{MS_EP_Context})-(\ref{TS_LLF_GM_Context}). \\
		$\mathbf{Augmented \; state \; extraction}$: \\
		if $r_k < 0.5$ no target is detected, \\
		otherwise do \\
		\qquad $\hat{c}_k = \arg \max_{c \in {\cal C}} \gamma_k(c)$ \\
		\qquad $\hat{m}_k = \arg \max_{m \in {\cal M}_{\hat{c}_k}} \beta^i_k(m|\hat{c}_k) $ \\
		\qquad Extract $\hat{x}_k$ from $s_k(\cdot|\hat{c}_k, \hat{m}_k)$ according to either \\ \qquad MAP or MMSE criterion \\
		\qquad Set $\hat{\mathbf{x}}_{k} = \left[ (\hat{x}_k)^T, \hat{c}_k, \hat{m}_k \right]^T$ \\
		end \\
		$\mathbf{Output}$: $f_{k}=\lbrace r_k, \gamma_k, \beta_k, s_k \rbrace$ and $\hat{\mathbf{x}}_{k}$ \\
		\bottomrule 
	\end{tabular} \label{MS_JDTC_MM_BF}
\end{table}

\section{Distributed JDTC-BF algorithm}

This section deals with the distributed setting wherein each sensor node $i \in {\cal N}$ computes a local posterior $f^i = \{ r^i, \gamma^i, \beta^i, s^i \}$ and fuses it with those of the
in-neighbors.
The idea is to approximate the global posterior
$f ({\cal X}_k|\bigcup\nolimits_{i \in \mathcal{N}} {\mathcal{Z}_{1:k}^i})$ in each node via repeated fusion (consensus) iterations with the neighboring nodes.
In particular, the
GCI fusion rule \cite{mahler2000optimal} is adopted by which the fused density is nothing but the geometrical average of the fusing ones, i.e.
\begin{align}
\overline{f}(\mathcal{X}) = \dfrac{\prod\limits_{i \in \mathcal{N}} {\left[  f^i(\mathcal{X}) \right]^{\omega^i}}}{ \displaystyle{\int} \prod\limits_{i \in \mathcal{N}} {\left[  f^i(\mathcal{X}) \right]^{\omega^i}} d\mathcal{X}} , \label{GCI_rule}
\end{align}
with suitably chosen fusion weights $\omega^i \in (0,1)$ such that $\sum_{i \in {\cal N}} \omega^i = 1$.
GCI fusion of Bernoulli RFS densities $f^i = \{ r^i, s^i \}$ into 
$\bar{f}=\lbrace \bar{r},\bar{s} \rbrace$ 
can be found in \cite[Eqs. (24)-(25)]{guldogan2014consensus}.
The next result provides an extension to the JDTC case with augmented Bernoulli RFS densities consisting of EP, CPMF, MPMFs and SPDFs
\begin{proposition}
	Given local Bernoulli RFS densities
	$f^i=\lbrace r^i,\gamma^i, \beta^i, s^i \rbrace $ and fusion weights $\omega^i$ for any $i \in \mathcal{N}$, the GCI-fused density in (\ref{GCI_rule})
	turns out to be a Bernoulli RFS density
	$\overline f =\lbrace \overline r, \overline \beta, \overline \gamma , \overline s\rbrace $
	given by
	\begin{align}
	\bar{r}&=\frac{\tilde{r}\sum\limits_{c \in {\cal C}}\tilde{\gamma}(c) \sum\limits_{m \in  \mathcal{M}_c} \tilde{\beta}(m|c)\int \tilde{s}(x|c,m) dx}
	{\tilde{\zeta}+\tilde{r}\sum\limits_{c \in {\cal C}}\tilde{\gamma}(c) \sum\limits_{m \in  \mathcal{M}_c} \tilde{\beta}(m|c)\int \tilde{s}(x|c,m)  dx},	\label{FUSION_RULE_EP}	\\
	\bar{\gamma}(c)&= \frac{\tilde{\gamma}(c) \sum\limits_{m \in \mathcal{M}_c}  \tilde{\beta}(m|c)  \int \tilde{s}(x|c,m) d x}{\sum\limits_{c \in {\cal C}}  \tilde{\gamma}(c)  \sum\limits_{m \in \mathcal{M}_c} \tilde{\beta}(m|c)  \int \tilde{s}(x|c,m) d x},  \\
	\bar{\beta}(m|c)&=\frac{\tilde{\beta}(m|c) \int \tilde{s}(x|c,m) d x}{\sum\limits_{m \in \mathcal{M}_c}  \tilde{\beta}(m|c)  \int \tilde{s}(x|c,m) d x}, \\
	\bar{s}(x|c,m)&=\frac{\tilde{s}(x|c,m)}{\int \tilde{s}(x|c,m) \,d x}, \label{fusion_SPDF}
	\end{align}
	with
	\begin{align}
	\tilde{s}(x|c,m)&=\prod\limits_{i \in \mathcal{N}}[ s^i(x|c,m) ]^{\omega^i}, \tilde{\beta}(m|c)=\prod\limits_{i \in \mathcal{N}}[ \beta^i(m|c) ]^{\omega^i}, \\
	\tilde{\gamma}(c)&=\prod\limits_{i \in \mathcal{N}}[ \gamma^i(c) ]^{\omega^i}, 
	\tilde{\zeta}=\prod\limits_{i \in \mathcal{N}}(1-r^i)^{\omega^i}, \tilde{r}=\prod\limits_{i \in \mathcal{N}}( r^i )^{\omega^i}. \label{zeta}
	\end{align}
	\label{prop3}
	\end{proposition}
\textit{Proof:} see Appendix $C$. \qed \\

For the sake of scalability, the global fusion (\ref{GCI_rule}) over the whole network $\mathcal{N}$ is actually replaced by a sequence of $L \geq 1$ fusion (consensus) steps over the subnetwork
${\cal N}^i$ containing node $i$ and its in-neighbors, i.e.  all nodes $j \neq i$ from which node $i$ has received data.
More precisely, (\ref{GCI_rule}) is replaced by the following iterative consensus procedure carried out in each node $i$:
\begin{equation}
\begin{array}{l}
\mbox{for } l=1, \dots, L \vspace{2mm} \\
f^i_{l} ( \mathcal{X} ) = \dfrac{\prod\limits_{j \in \mathcal{N}^i} {\left[  f_{l-1}^j(\mathcal{X}) \right]^{\omega^{i,j}}}}{ \displaystyle{\int} \prod\limits_{j \in \mathcal{N}^i} {\left[  f_{l-1}^j(\mathcal{X}) \right]^{\omega^{i,j}}} d\mathcal{X}} 
\end{array}
\label{consensus}
\end{equation}
initialized from $f^i_0(\mathcal{X}) = f^i(\mathcal{X})$ and with consensus weights $\omega^{i,j} > 0$, satisfying $\sum_{j \in {\cal N}^i} \omega^{i,j} = 1$, possibly selected so as to 
ensure that $f_l^i (\mathcal{X})$ converges to $\overline{f}(\mathcal{X})$ as $l \rightarrow \infty$. 
The resulting D-JDTC-BF algorithm is summarized in Table \ref{DJDTC_MM_BF}.

\begin{table}[htbp] 
	\caption{\label{tab:test}} 
	\begin{tabular}{lcl} 
		\toprule 
		$\mathbf{Algorithm}$: D-JDTC-BF (Node $i$, Time $k$) \\ 
		\midrule 
		$\mathbf{Input}$: $\mathcal{Z}^i_k, f^{i}_{k-1}=\lbrace r^i_{k-1},\gamma^i_{k-1},\beta^i_{k-1},s^i_{k-1}\rbrace$ \\
		$\mathbf{Local \; filtering}$: \\
		Carry out local filtering with prediction and update \\
		steps of  JDTC-BF  to 
		get the local Bernoulli density \\
		$f^{i}_{k,0}=\lbrace r^{i}_{k,0},\gamma^{i}_{k,0},\beta^{i}_{k,0} ,s^{i}_{k,0}\rbrace $; \\
		$\mathbf{GCI} \; \mathbf{fusion}$: \\
		for $l=1, \cdots , L$ do \\
		\qquad 1. Receive data from in-neighbors $j \in \mathcal{N}^i \backslash \{i \}$ \\
		 \qquad \qquad to get $f^j_{k,l-1} = \{ r^j_{k,l-1}, \gamma^j_{k,l-1},  \beta^j_{k,l-1}, s^j_{k,l-1} \}$  \\
		\qquad 2. Fuse $\{ f^j_{k,l-1} \}_{j \in \mathcal{N}^i}$ with weights $\omega^{i,j}$
		via \\ 
		\qquad \qquad (\ref{FUSION_RULE_EP})-(\ref{zeta}) to get $f^j_{k,l} = \{ r^j_{k,l}, \gamma^j_{k,l},  \beta^j_{k,l}, s^j_{k,l} \}$;\\
		end\\
		Set $f_k^i = \{ r_k^i, \gamma_k^i, \beta_k^i, s_k^i \} =  \{ r_{k,L}^i, \gamma_{k,L}^i, \beta_{k,L}^i, s_{k,L}^i \}$ \\
		$\mathbf{Augmented \; state \; extraction}$: \\
		if $r_k^i < 0.5$ no target is detected, \\
		otherwise do \\
		\qquad $\hat{c}_k^i = \arg \max_{c \in {\cal C}} \gamma_k^i(c)$ \\
		\qquad $\hat{m}_k^i = \arg \max_{m \in {\cal M}_{\hat{c}_k^i}} \beta^i_k(m|\hat{c}_k^i) $ \\
		\qquad Extract $\hat{x}_k^i$ from $s_k^i(\cdot|\hat{c}_k^i, \hat{m}_k^i)$ according to either \\ \qquad MAP or MMSE criterion \\
		\qquad Set $\hat{\mathbf{x}}_{k}^i = \left[ (\hat{x}_k^i)^T, \hat{c}_k^i, \hat{m}_k^i \right]^T$ \\
		end \\
$\mathbf{Output}$: $f^{i}_{k}=\lbrace r^i_k,\gamma^i_k,\beta^i_k , s^i_k\rbrace$ and $\hat{\mathbf{x}}^i_{k}$ \\ 
		
		\bottomrule 
	\end{tabular} \label{DJDTC_MM_BF}
\end{table}

\begin{remark}
	It is worth pointing out that the D-JDTC-BF approach of this section significantly extends previous work in
	\cite{battistelli2015consensus} on distributed multiple-model Bayesian tracking of a maneuvering target.
	In fact, D-JDTC-BF allows to perform also target detection and classification, besides tracking, and considers the presence of clutter as well as target appearance/disappearance.
\end{remark}

\section{Gaussian-mixture implementation of JDTC-BF}

In this section,
the JDTC-BF is implemented by utilizing the Gaussian mixture (GM) approach \cite{vo2007analytic}.
For the subsequent developments, the target motion is modeled by a mode-dependent state equation of the form
\begin{equation}
\begin{array}{rcl}
{x}_{k} & = & f({x}_{k-1},m_{k})+ w_k  \vspace{2mm} \\
w_k & \sim & {\cal G}(\cdot; 0, Q(m_k))
\end{array}
\label{motion}
\end{equation}
where $w_k$ is a Gaussian process noise with zero mean and mode-dependent covariance $Q(m_k)$.
Accordingly, the kinematic state transition density is
\begin{equation}
\varphi (x_+ | x , m_+) =  {\cal G}(x_+; f(x,m_+) , Q(m_+)) \, .
\end{equation}
Further, each sensor $i$ is modeled by a measurement equation of the form
\begin{equation}
\begin{array}{rcl}
z^i_k &  = & h^i(x_k)+ v_k^i \vspace{2mm} \\
v_k^i & \sim & {\cal G} \left( \cdot; 0, R^i \right)
\end{array}
\label{measurement}
\end{equation}
where $v_k^i $ is a Gaussian measurement noise with zero mean and  covariance $R^i$.
The measurement noises $v_k^i $ of different sensors are assumed mutually independent and independent of the process noise $w_k$.
Accordingly, the likelihood function associated to the $i^{th}$ sensor measurement model in (\ref{target-meas}) is given by
\begin{equation}
 \ell^i(z|x) = {\cal G} \left( z; h^i(x), R^i \right).
 \label{lik}
\end{equation}

Hereafter, in order to approximate the likelihood function of nonlinear measurement (\ref{measurement}),
we follow the EKF approach and linearize the function $h^i$ based on the Taylor expansion at $x_0$
\begin{equation}
	h^i(x) \cong h^i(x_0)+H^i(x_0)(x-x_0),
\end{equation}
where $H^i$ is the Jacobian matrix. Similarly, the function $f$ is linearized as
\begin{equation}
	f(x,m) \cong f(x_0,m)+F(x_0,m)(x-x_0) 
\end{equation}
where $F(x_0,m)$ is the Jacobian matrix with respect to $x$ for a given $m$.


In the proposed implementation scheme,
the GM is utilized to approximately represent SPDFs of the Bernoulli density.
In particular,
for a Bernoulli density $f=\lbrace r,\gamma,\beta,s \rbrace$,
a GM with $J{(c,m)}$ Gaussian components (GCs) is employed for the SPDF 
\begin{align}
s(x|c,m)= \sum_{j=1}^{J(c,m)} \alpha_j(c,m) \cdot \mathcal{G}(x;\mu_j(c,m),P_j(c,m)),
\end{align}
where $\alpha_j(c,m),
\mu_j(c,m)$ and $P_j(c,m)$ are the weight, 
mean and covariance of $j$-th GC conditioned on the class $c \in \mathcal{C}$ and mode $m \in \mathcal{M}_c$.
Thus,
the SPDF can be more compactly rewritten as $s=\lbrace \alpha_j,\mu_j,P_j \rbrace^{J(c,m)}_{j=1}$.

\subsection{Prediction}

At time $k-1$, given 
the prior augmented Bernoulli density $f_{k-1} = \lbrace r_{k-1},\gamma_{k-1},\beta_{k-1},s_{k-1} \rbrace$ with class\&mode-conditioned SPDFs represented in GM form
$$ 
\begin{array}{l}
s_{k-1}(x|c,m)  = \displaystyle{\sum_{j=1}^{J_{k-1}(c,m)}} \alpha_{k-1,j}(c,m)  \vspace{2mm} \\ \times \, {\cal G} \left( x; \mu_{k-1,j}(c,m),P_{k-1,j}(c,m) \right)
\end{array}
$$
 and the augmented birth Bernoulli density $f_B
=\lbrace p_B,\gamma_B,\beta_B, s_B \rbrace$ with SPDFs represented in GM form 
$$ 
\begin{array}{l}
s_{B}(x|c,m) =  \displaystyle{\sum_{j=1}^{J_{B}(c,m)}} \alpha_{B,j}(c,m) {\cal G} \left( x; \mu_{B,j}(c,m), P_{B,j}(c,m) \right),
\end{array}
$$
then the class$\&$mode-conditioned SPDF of the augmented predicted density $f_{k|k-1}=\lbrace r_{k|k-1},\beta_{k|k-1},\gamma_{k|k-1},s_{k|k-1} \rbrace $ via (\ref{pred1})-(\ref{pred4})
is given by 
\begin{align}
&s_{k|k-1}(x|c,m) \nonumber\\
&= \frac{p_B(1-r_{k-1})\gamma_B(c)\beta_B(m|c)}{r_{k|k-1}\gamma_{k|k-1}(c)\beta_{k|k-1}(m|c)} \nonumber\\
&\times \sum\limits^{J_B(c,m)}_{j=1}\alpha_{B,j}(c,m) \, \mathcal{G}(x;\mu_{B,j}(c,m),P_{B,j}(c,m)) \nonumber \\
&+\frac{p_S \, r_{k-1}\gamma_{k-1}(c) }{r_{k|k-1}\gamma_{k|k-1}(c) \, \beta_{k|k-1}(m|c)} \sum\limits_{m' \in \mathcal{M}_c}\pi_c(m|m')\beta_{k-1}(m'|c)   \nonumber \\
& \times \sum\limits^{J_{k-1}(c,m')}_{j=1 } \alpha_{k-1,j}(c,m')   \nonumber \\
&\times
\mathcal{G}({x;\mu^S_{k|k-1,j}(c,m,m'),P^S_{k|k-1,j}(c,m,m')}), \label{SPDF_predict}
\end{align} 
where
\begin{align}
\mu^S_{k|k-1,j}(c,m,m')&= f(\mu_{k-1,j}(c,m'),m)  \label{pre_mean} \\
P^S_{k|k-1,j}(c,m,m')&=	F_{k,j} P_{k-1,j}(c,m') F_{k,j}^{\top}+Q_k{(m)} \label{pre_cov} \\
F_{k,j} & = \frac{\partial f}{\partial x}(\mu_{k-1,j}(c,m'),m) .
\end{align}
Notice that the predicted densities $s_{k|k-1}(x|c,m)$ in (\ref{SPDF_predict}) are still in GM form, but with an increased number of GCs given by
\begin{equation}
J_{k|k-1}(c,m)= J_B(c,m) + \sum\limits_{m' \in \mathcal{M}_c}J_{k-1}(c,m'),
\end{equation}
and can, therefore, be rearranged as
\begin{equation}
\begin{array}{l}
s_{k|k-1}(x|c,m)  =  \sum_{j=1}^{J_{k|k-1}(c,m)} \alpha_{k|k-1,j}(c,m)  \vspace{2mm} \\ 
\times {\cal G} \left( x; \mu_{k|k-1,j}(c,m),P_{k|k-1,j}(c,m) \right)
\end{array}
\label{GM-SPDF-pred}
\end{equation}
for appropriate weights, means and covariances of the GCs.

\subsection{Single-sensor update of local JDTC-BF}
Let us now consider single-sensor update with the local measurement set $\mathcal{Z}_k = \mathcal{Z}_k^i$, omitting for the sake of simplicity superscript $i$.
Starting from the predicted density $f_{k|k-1}=\lbrace r_{k|k-1},\gamma_{k|k-1},\beta_{k|k-1},s_{k|k-1} \rbrace$,
we can therefore apply (\ref{MS_EP_Context})-(\ref{MS_SPDF_GM_Context}) to get $f_k  = \{ r_k, \gamma_k, \beta_k, s_k \}$.
Exploiting the GM form (\ref{GM-SPDF-pred}) of $s_{k|k-1}(x|c,m)$ and (\ref{lik}), the likelihoods (\ref{TM_LLF_GM_Context})
take the form
\begin{align}
\ell(m|c)&= [1-p_D(c)]   \nonumber \\
&+ p_D(c) \sum\limits_{z \in \mathcal{Z}_k}\sum\limits_{j=1}^{J_{k|k-1}(c,m)} {\alpha}_{k,j}(c,m,z)  \label{MP_likelifunc} .
\end{align}
Accordingly, the updated SPDFs in (\ref{MS_SPDF_GM_Context}) become as follows
\begin{align}
s_k(x|c,m) 
&=\frac{1}{\Lambda} \cdot\Bigg [ \left[ 1-p_{D}(x,c) \right] \sum_{j=1}^{J_{k|k-1}(c,m)}\alpha_{k|k-1,j}(c,m) \nonumber \\
&\times \mathcal{G}(x;\mu_{k|k-1,j}(c,m),P_{k|k-1,j}(c,m)) \nonumber \\
&+  p_{D}(c) \sum\limits_{z \in \mathcal{Z}_k}\sum_{j=1}^{J_{k|k-1}(c,m)}{\alpha}_{k,j}(c,m,z) \nonumber \\
&\times \mathcal{G}(x;\mu_{k,j}(c,m,z),P_{k,j}(c,m))\Bigg ]
\label{update_SPDF} \\
\Lambda&=[1-p_D(c)]   \nonumber \\
&+ p_D(c) \sum\limits_{z \in \mathcal{Z}_k}\sum\limits_{j=1}^{J_{k|k-1}(c,m)} {\alpha}_{k,j}(c,m,z)\\
{\alpha}_{k,j}(c,m,z)&=\dfrac{q(z)}{\kappa(z)}{ \alpha}_{k|k-1,j}(c,m)
\\
q(z)&=\mathcal{G}(z; h \left( \mu_{k|k-1,j}(c,m)),S_{k,j} \right) \\
{\mu}_{k,j}(c,m,z) &={\mu}_{k|k-1,j}(c,m)+K_{k,j} \nonumber \\ &\times \left( z- h \left( \mu_{k|k-1,j}(c,m) \right) \right) \label{update_mean} \\
H_{k,j} & = \frac{\partial h}{\partial x} \left( \mu_{k|k-1,j}(c,m) \right) \\
{P}_{k,j}(c,m) &=(I-K_{k,j}H_{k,j}){P}_{k|k-1,j}(c,m) \label{update_cov} \\
K_{k,j} &={P}_{k|k-1,j}(c,m) H^{\top}_{k,j} S_{k,j}^{-1} \label{update_gate} \\
S_{k,j} &=R +H_{k,j}{P}_{k|k-1,j}(c,m)  H^{\top}_{k,j} \, . \label{update_gate_cov} 
\end{align} 

Notice that (\ref{update_SPDF}) is in GM form, as required for the subsequent steps, with number of GCs
\begin{equation}
 J_k(c,m) = \left( | {\cal Z}_k | + 1 \right) J_{k|k-1}(c,m) 
\end{equation}
increased by a factor equal to the number of measurement plus one.

\begin{remark}
	JDTC aims to jointly estimate target state, class and mode.
	In \cite{yang2015joint},
	the proposed method solves the problem by exploiting a \textit{sequential Monte Carlo} (SMC) implementation of the Bernoulli filter.
	Since typically GM representation of a PDF is more parsimonious than SMC representation and also because JDTC involves multiple SPDFs for all class-mode pairs
	$(c,m)$, the GM approach seems by far preferable especially in the distributed case wherein posteriors are transmitted and received by each sensor node.
	Unfortunately, however, the number of GCs increases at each Bayesian step (prediction, update or fusion) 
   so that suitable
   pruning and/or merging procedures (see \cite[Table $ \uppercase \expandafter {\romannumeral 2}$]{vo2006gaussian})  are needed to limit such a number.
\end{remark}

\subsection{Multi-sensor update of C-JDTC-BF}

Proposition $2$ provides the update equations based on Bayes-optimal C-JDTC-BF.
In the multi-sensor case, 
for the class$\&$mode-conditioned SPDF,
it is possible to implement centralized fusion by iterating single-sensor Bernoulli filter updates as follows \cite[Section $\uppercase \expandafter {\romannumeral 4}-B$]{vo2012multi}.

\begin{itemize}
	\item Starting from the prior $s^{(0)}_k= s_{k|k-1}$, first compute  
	$ s^{(1)}_{k}=\lbrace \alpha^{(1)}_{k,j},\mu^{(1)}_{j,k},P^{(1)}_{k,j} \rbrace^{J^{(1)}_{k}(c,m)}_{j=1}$ using measurements and parameters of sensor $1$ 
	according to the single-sensor Bernoulli filter update procedure of the previous subsection.
	\item Next, apply to the prior $s^{(1)}_k$ the same procedure, with measurements and parameters of sensor $2$, to get
	$ s^{(2)}_{k}=\lbrace \alpha^{(2)}_{k,j},\mu^{(2)}_{k,j},P^{(2)}_{k,j} \rbrace^{J^{(2)}_{k}(c,m)}_{j=1}$.
	\item Repeat the same step until all sensors have been considered and $s_k = s_k^{(| {\cal N}|)}$ is obtained.
\end{itemize}

The multi-sensor updated Bernoulli density $f_{k}
=\lbrace r_{k},\gamma_k,\beta_k,s_k \rbrace$ can be obtained by using (\ref{MS_EP_Context})-(\ref{MS_SPDF_GM_Context}).
Furthermore,
the updated SPDFs in (\ref{MS_SPDF_GM_Context}) are computed by using the above-mentioned iterations
\begin{align}
s_{k}(x|c,m)&=\Psi^{(|{\cal N}|)}_k \circ \cdots \circ \Psi^{(2)}_k \circ \Psi^{(1)}_k [s_{k|k-1}(x|c,m)]
\end{align}
where $\circ$ denotes composition,
and
\begin{align}
&\Psi^{(i)}_k [s_k^{(i-1)}(x|c,m)] \nonumber \\
&=\frac{1}{\Lambda^{(i)}} \Bigg [ \left[ 1-p^{(i)}_{D}(c) \right] \sum_{j=1}^{J^{(i-1)}_{k}(c,m)} \tilde{\alpha}^{(i-1)}_{k,j}(c,m)  \nonumber \\
&\times \mathcal{G}(x;\tilde{\mu}^{(i-1)}_{k,j}(c,m),\tilde{P}^{(i-1)}_{k,j}(c,m)) \nonumber \\
&+  p^{(i)}_{D}(c) \sum\limits_{z \in \mathcal{Z}^{i}_k}\sum_{j=1}^{J^{(i-1)}_{k}(c,m)}{\alpha}^{(i)}_{k,j}(c,m,z)  \nonumber \\
&\times \mathcal{G}(x;\mu^{(i)}_{k,j}(c,m,z),P^{(i)}_{k,j}(c,m))\Bigg ] \nonumber \\
&=\frac{1}{\Lambda^{(i)}} \Bigg [ \sum\limits^{J^{(i)}_k(c,m)}_{j=1} \tilde{\alpha}^{(i)}_{k,j}(c,m) \nonumber \\
&\times \mathcal{G}(x;\tilde{\mu}^{(i)}_{k,j}(c,m),\tilde{P}^{(i)}_{k,j}(c,m))  \Bigg ] \label{MS_update_SPDF} 
\end{align}
where $\tilde{\alpha}^{(0)}_{k,j}(c,m)={\alpha}_{k|k-1,j}(c,m),\tilde{\mu}^{(0)}_{k,j}(c,m)={\mu}_{k|k-1,j}(c,m),\tilde{P}^{(0)}_{k,j}(c,m)={P}_{k|k-1,j}(c,m)$,
and
\begin{align}
\Lambda^{(i)}&=\left[ 1-p^{(i)}_{D}(c) \right]  \nonumber \\
&+  p^i_{D}(c) \sum\limits_{z \in \mathcal{Z}^{(i)}_k}\sum_{j=1}^{J^{(i-1)}_{k}(c,m)}{\alpha}^{(i)}_{k,j}(c,m,z)  \\
J^{(i)}_k(c,m) &=J^{(i-1)}_{k}(c,m)+|\mathcal{Z}^i_k|J^{(i-1)}_{k}(c,m)\,\\
{\alpha}^{(i)}_{k,j}(c,m,z)
&=\dfrac{q^{(i)}(z)}{\kappa(z)}{ \tilde{\alpha}}^{(i-1)}_{k,j}(c,m) 
 \\
q^{(i)}(z)&= \mathcal{G} \left(z; h ( \tilde{\mu}^{(i-1)}_{k,j}(c,m)),S^{(i)}_{k,j} \right) \\
{\mu}^{(i)}_{k,j}(c,m,z) &=\tilde{\mu}^{(i-1)}_{k,j}(c,m)+K^{(i)}_{k,j} \nonumber \\ &\times \left( z- h \left( \tilde{\mu}^{(i-1)}_{k,j}(c,m) \right) \right)  \\
H^{(i)}_{k,j} & = \frac{\partial h}{\partial x} \left( \tilde{\mu}^{(i-1)}_{k,j}(c,m) \right) \\
{P}^{(i)}_{k,j}(c,m) &=(I-K^{(i)}_{k,j}H^{(i)}_{k,j})\tilde{P}^{(i-1)}_{k,j}(c,m)  \\
K^{(i)}_{k,j} &=\tilde{P}^{(i-1)}_{k,j}(c,m) (H^{(i)}_{k,j})^{\top} (S^{(i)}_{k,j})^{-1} \\
S^{(i)}_{k,j} &=R^{(i)} +H^{(i)}_{k,j}\tilde{P}^{(i-1)}_{k,j}(c,m)  (H^{(i)}_{k,j})^{\top} \, .
\end{align}

Moreover,
the likelihood function of class-based mode in (\ref{TM_LLF_GM_Context}) equals  the normalization value of the class$\&$mode-condtioned SPDF,
thus,
$\ell(m|c)=\Lambda^{(|\mathcal{N}|)}$ can be obtained by performing iterations until all sensors have been considered.

At the end of the above multi-sensor update procedure, the resulting posterior SPDFs $s_k(x|c,m)$ preserve the GM form with a number of GCs given by
\begin{equation}
J_{k}(c,m) =  \left[ \displaystyle{\prod_{i \in {\cal N}}} \left( | {\cal Z}_k^i |+ 1 \right)  \right] \, J_{k|k-1}(c,m) .
\label{GCN}
\end{equation}

\subsection{Fusion of D-JDTC-BF}
This subsection concerns GM implementation of the fusion stage of D-JDTC-BF.
For the sake of simplicity,
we only consider pairwise fusion of two augmented Bernoulli densities $f^i=\lbrace r^i,\gamma^{i},\beta^{i},s^i \rbrace$, for $i=1,2$, with SPDFs in GM form
\begin{align}
s^{i}(x|c,m)=\sum\limits^{J^i(c,m)}_{j=1} \alpha^{i}_j(c,m) \mathcal{G}(x;\mu^{i}_j(c,m),P^{i}_j(c,m) ) \,\, i=1,2. \label{GM_JSPDF}
\end{align}
As well known,  the weighted geometric average of GMs
\begin{equation}
\tilde{s}(x|c,m) = \left[ s^1(x|c,m) \right]^\omega \, \left[ s^2(x|c,m) \right]^{1-\omega},
\label{WGA}
\end{equation}
due to exponentiation by the fractional exponents $\omega$ and $1-\omega$,
is no longer a GM.
However, there exist reasonable approximations of (\ref{WGA}) in GM form like, e.g., \cite{battistelli2013consensus}
\begin{equation}
\begin{array}{rcl}
\tilde{s}(x|c,m) & \cong & \displaystyle{\sum_{j_1=1}^{J^1(c,m)}}  \, 
\displaystyle{\sum_{j_2=1}^{J^2(c,m)}}   \alpha_{j_1,j_2}(c,m)  \vspace{2mm} \\
& & \times \,
{\cal G} \left( x; \mu_{j_1,j_2}(c,m), P_{j_1,j_2}(c,m)  \right) \vspace{2mm} \\
\alpha_{j_1.j_2}(c,m) & = & (\alpha_{j_1}^1)^\omega \,  (\alpha_{j_2}^2)^{1-\omega} \vspace{2mm} \\
 & &   \varepsilon(\omega,P_{j_1}^1)   \, \varepsilon(1-\omega,P_{j_2}^2) \vspace{2mm} \\
& & {\cal G} \left( \mu_{j_1}^1 - \mu_{j_2}^2; 0, \frac{P_{j_1}^1}{\omega} + 
 \frac{P_{j_2}^2}{1-\omega} \right) \vspace{2mm} \\
 P_{j_1,j_2}(c,m) & = & \left[ \omega (P_{j_1}^{1})^{-1} + (1-\omega) (P_{j_2}^{2})^{-1} \right]^{-1} \vspace{2mm} \\
\mu_{j_1,j_2}(c,m) & = &P_{j_1,j_2}(c,m) \left[ \omega (P_{j_1}^1)^{-1} + (1-\omega) (P_{j_2}^2)^{-1} \right] \vspace{2mm} \\
\varepsilon(\omega,P) & = & \sqrt{\det(2\pi P \omega^{-1}) \, det(2\pi P)^{-\omega}}
\end{array}
\label{appWGA}
\end{equation}
where, for the sake of brevity, the arguments $c$ and $m$ of $\alpha_j^i, \mu_j^i, P_j^i$ have been omitted.
An alternative approximation, more appropriate than (\ref{appWGA}) whenever there are closely-spaced GCs, can be found in \cite{orguner}.
It is worth to point out that both approximations in \cite{battistelli2013consensus,orguner} produce a GM with number of GCs given by
\begin{equation}
 J(c,m) = J^1(c,m) \, J^2(c,m)\, .
\end{equation}
Then, according to Proposition \ref{prop3}, fused EP, MPMFs, CPMFs and SPDFs are obtained from (\ref{FUSION_RULE_EP})-(\ref{fusion_SPDF}) by setting
\begin{eqnarray}
\int \tilde{s}(x|c,m) dx & = & \displaystyle{\sum_{j_1=1}^{J^1(c,m)}} \, \displaystyle{\sum_{j_2=1}^{J^2(c,m)}}   \alpha_{j_1,j_2}(c,m) \\
\tilde{r} & = & (r^1)^\omega \, (r^2)^{1-\omega} \\
\tilde{\gamma}(c) & = & \left[ \gamma^1(c) \right]^\omega \,  \left[ \gamma^2(c) \right]^{1-\omega} \\
\tilde{\beta}(m|c) & = & \left[ \beta^1(m|c) \right]^\omega  \, \left[ \beta^2(m|c) \right]^{1-\omega} \\
\tilde{\zeta} & = & (1 - r^1)^\omega \, (1-r^2)^{1-\omega} . 
\end{eqnarray}
Whenever fusion involves more than two sensor nodes, it is anyway possible to split it, in several ways, into a sequence of pairwise fusion steps to be performed as indicated above.
Whatever is the adopted sequence, fusion over ${\cal N}$ of Proposition  \ref{prop3} generates GM SPDFs $\overline{s}(x|c,m)$ with 
\begin{equation}
\overline{J}(c,m) = \displaystyle{\prod_{i \in {\cal N}}} J^i(c,m)
\end{equation}
GCs, where $J^i(c,m)$ is the number of GCs of $s^i(x|c,m)$.

\section{Simulation Results}

Assume that the target can belong to three different classes with corresponding 
mode sets $\mathcal{M}_1=\lbrace m_1 \rbrace$,
$\mathcal{M}_2=\lbrace m_1, m_2, m_3\rbrace$
and $\mathcal{M}_3=\lbrace m_1,m_4, m_5\rbrace$.
The target kinematic state at time $k$ is defined as  $x_k=[\xi_k,\dot{\xi}_k,\eta_k,\dot{\eta}_k] ^\top$,
with Cartesian coordinates of position $\xi_k,\eta_k$ and, respectively, velocity $\dot{\xi}_k,\dot{\eta}_k$.
For each kinematic mode,
the target motion is modeled by (\ref{motion}) with
$$
 f(x,m) \, = F(m) x.
$$
There are five possible modes and their corresponding state transition $F(m)$ and process noise covariance $Q(m)$ matrices are as follows.

\emph{Mode} $m_{1}$:
\begin{align}   
F(m_{1}) &= \left[ {\begin{array}{*{20}{l}}
	1&T&0&0\\
	0&1&0&0\\
	0&0&1&T\\
	0&0&0&1
	\end{array}} \right],   \\
Q(m_{1}) &=\sigma(m_1) \left[ {\begin{array}{*{20}{l}}
	{\frac{{T^3}}{3}}&{\frac{{{T^2}}}{2}}&0&0\\
	{\frac{{{T^2}}}{2}}&T&0&0\\
	0&0&{\frac{T^3}{3}}&{\frac{{{T^2}}}{2}}\\
	0&0&{\frac{{{T^2}}}{2}}&T^2
	\end{array}} \right], \label{PNM_1}
\end{align}
where $T=1 [s]$ denotes the sampling interval
and $\sigma(m_1)=1[m/s^2]$.

\emph{Mode} $m_{2}$:
\begin{align}   
F(m_{2}) &= \left[ {\begin{array}{*{20}{l}}
	1&sin(\omega T)/\omega&0& (cos(\omega T)-1)/\omega\\
	0&cos(\omega T)&0&-sin(\omega T)\\
	0&(1-cos(\omega T))/\omega &1&sin(\omega T)/\omega\\
	0&sin(\omega T)&0&cos(\omega T)
	\end{array}} \right],  \label{TM_2} \\
Q(m_{2}) &= \sigma(m_2)\left[ {\begin{array}{*{20}{l}}
	{\frac{{3{T^4}}}{4}}&{\frac{{{T^3}}}{2}}&0&0\\
	{\frac{{{T^3}}}{2}}&T^2&0&0\\
	0&0&{\frac{3{{T^4}}}{4}}&{\frac{{{T^3}}}{2}}\\
	0&0&{\frac{{{T^3}}}{2}}&T^2
	\end{array}} \right] \label{Process_noise_cov}
\end{align}
where $\omega=-0.1 [rad/s]$ 
and $\sigma(m_2)=1.4[m/s^2]$.

\emph{Mode} $m_{3}$: same as  mode $m_2$ with $\omega=0.15 [rad/s]$ and $\sigma(m_3)=1.4[m/s^2]$.

\emph{Mode} $m_{4}$: same as  mode $m_2$ with $\omega=1 [rad/s]$ and $\sigma(m_4)=1.4[m/s^2]$.

\emph{Mode} $m_{5}$: same as  mode $m_2$ with $\omega=-1 [rad/s]$ and $\sigma(m_5)=1.4[m/s^2]$.

Transition probability matrices for classes 2 and 3 are  
\begin{align}
\Pi = \left[ {\begin{array}{*{20}{l}}
	0.8&0.1&0.1\\
	0.1&0.8&0.1\\
	0.1&0.1&0.8
	\end{array}} \right].
\end{align}
while class 1 has only one mode, i.e. $\Pi = 1$.
\begin{figure}[h]
	\centering
	\includegraphics[width=0.52\textwidth]{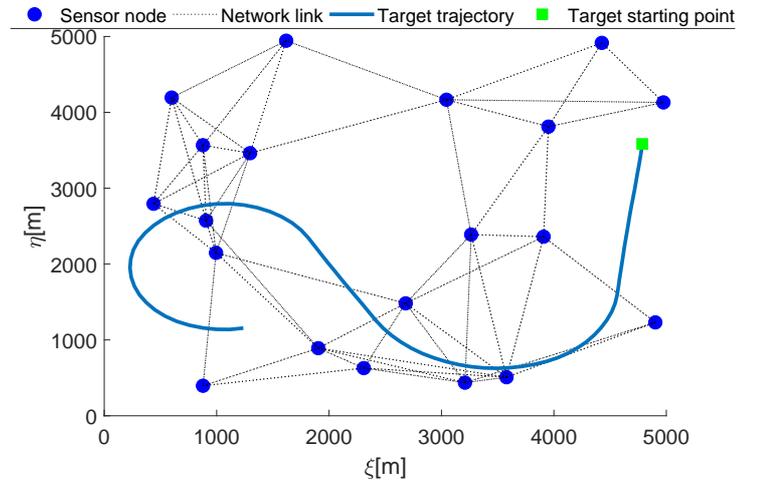}\\
	\caption{Sensor network and target track  }
	\label{fig1}
\end{figure}

This simulation duration is $100 [s]$ for each experiment.
The maneuvring target, belonging to class 2, appears at time  $t_a= 6[s]$ and disappears at $t_d=90[s]$
in the surveillance region of angle extension $[0,\pi/2]$ and range extension $ [0,5000 \sqrt{2}]$.
The initial target state is: 
\begin{align}
x=[4786[m], -8.3[m/s], 3584[m], -100.9[m/s]]^{\top}.
\end{align}

\begin{table*}[]
	
	\scriptsize
	
	\centering
	
	\caption{Target modes and classes over time}
	
	\label{Tab03}
	
	\begin{tabular}{cccccccccc}
		
		\toprule
		
		\multirow{1}{*}{Time} & \multicolumn{1}{c}{1-5[s]} & \multicolumn{1}{c}{6-25[s]} 
		& \multicolumn{1}{c}{26-50[s]}& \multicolumn{1}{c}{51-60[s]}& \multicolumn{1}{c}{61-90[s]}& \multicolumn{1}{c}{91-100[s]} \\

		\midrule
		
		True Mode          &Disappearance                     & Mode 1                 &  Mode 2             & Mode 1   & Mode 3     & Disappearance            \\
		Possible class           &Disappearance                     & Class 1,Class 2,class 3                & Class 2              & Class 1,Class 2,class 3  & Class 2        & Disappearance        \\
		True class            &Disappearance                      & Class 2                 & Class 2      & Class 2      & Class 2           & Disappearance    \\
		
		\bottomrule
		
	\end{tabular} \label{M_C_of_target}
	
\end{table*}
Its trajectory (see Fig. \ref{fig1}) is a straight line with constant velocity between $6[s]$ and $25[s]$, followed by a 
a clockwise turn ($\omega=-0.10[rad/s]$) between $26[s]$ and $50[s]$,
another straight line with constant velocity between $51[s]$ and $60[s]$, and a final counterclockwise
turn ($\omega=0.15[rad/s]$) between $61[s]$ and $90[s]$.
Target modes over time are reported in Table \ref{M_C_of_target}.

A total of $|{\cal N}|=20$ sensors are deployed over the surveillance area as shown in Fig. \ref{fig1}.
Each sensor $i \in {\cal N}$, of known position $(\xi^i,\eta^i)$, provides a range measurement according to (\ref{target-meas}) with measurement function
\begin{align}
h^i({x})=\sqrt{(\xi-\xi^i)^2+(\eta-\eta^i)^2}, 
\end{align}
uniform detection probability $p_D^i=0.95$ and measurement noise variance  $R^i=25[m^2]$.
Clutter is generated as a Poisson RFS with PHD 
$\kappa(z)=\lambda  u(z)$, with expected number of clutter points $\lambda = 5 $
and uniform spatial PDF $u(\cdot)$  over the surveillance region.

The JDTC-BF has been tuned with survival probability
$p_S=0.98$ and birth probability $p_B=0.2$.
Further, for target birth we assumed uniform distribution for class and mode as well as class-and-mode independent Gaussian distribution for the state.
Specifically, 
the target birth PDF has been taken as $b(x,c,m) = \gamma_B(c) \beta_B(m|c) s_B(x)$ with
\begin{equation}
\begin{array}{rcc}
\gamma_B(c)         & = & \dfrac{1}{| {\cal C} |} \vspace{2mm} \\
\beta_B(m|c) & = & \dfrac{1}{| {\cal M}_c|} \vspace{2mm} \\
s_B(x) & = & {\cal G} \left( x; m_B, P_B \right)
\end{array}
\end{equation}
where $m_B=[4780[m], -8[m/s], 3590[m], -100[m/s]]^{\top}$ and $P_B=diag([100[m^2], 100[m^2/{s^2} ], 100[m^2], 100[m^2/ {s^2}]])$.
For D-JDTC-BF, the number of consensus steps has been set to $L=3$.
In particular,
the pruning and merging thresholds for GM are set to $T_p=1 \times 10^{-15}$ and $T_m=20$,
respectively.
Moreover,
the maximum number of GCs is fixed to 6.

In the simulation experiments, performance will be evaluated in terms of 
four performance indicators averaged over $100$ independent Monte Carlo trials:
\textit{optimal sub-pattern assignment} (OSPA) error \cite{schuhmacher2008consistent} (with order $p=1$ and cutoff $c=150[m]$),
existence probability,
mode PMF,
and classification probability. 

\begin{figure}[h]
	\centering
	\includegraphics[width=0.52\textwidth]{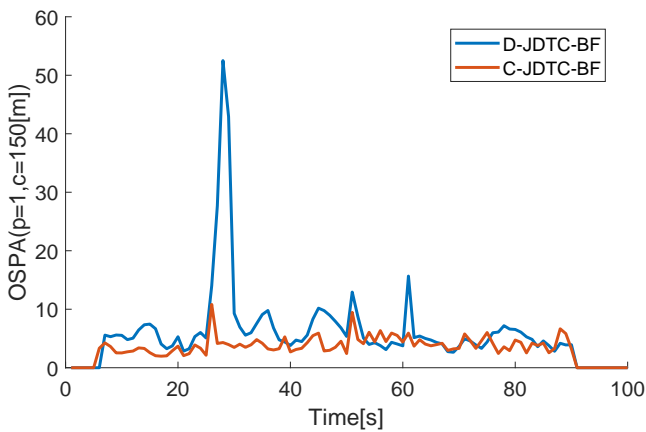}\\
	\caption{OSPA error}
	\label{fig2}
\end{figure}

The OSPA error is plotted in Fig. \ref{fig2} where
it can be seen how C-JDTC-BF clearly provides lower OSPA than D-JDTC-BF as well as a smoother behavior during class and/or mode switches.
Specifically,
D-JDTC-BF exhibits OSPA peaks after each class and/or mode switch:
a first peak in the time interval $26-30[s]$ is due to a simultaneous class and mode switch, while the subsequent  two peaks (at time $51[s]$ and $61[s]$) 
are caused by mode switches.

\begin{figure}[h]
	\centering
	\includegraphics[width=0.52\textwidth]{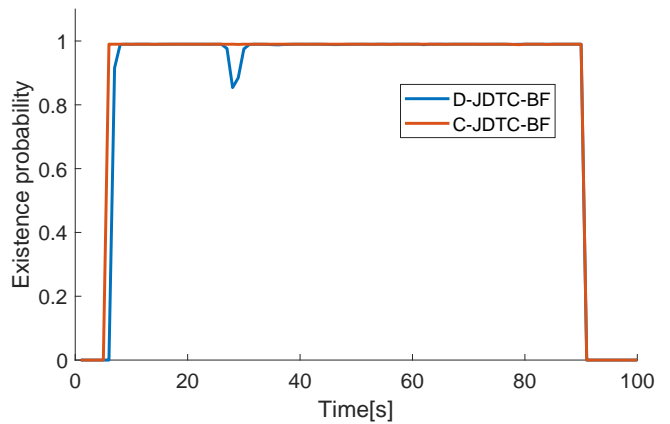}\\
	\caption{Existence probability}
	\label{fig3}
\end{figure}
As well known, the OSPA error simultaneously captures
detection and tracking performance. A more clear-cut assessment of detection performance 
is provided by Fig. \ref{fig3}, plotting the estimated existence probability. 
It can be seen that C-JDTC-BF has better detection capability than  D-JDTC-BF which exhibits a decrease of the existence probability
in the time interval $26-30[s]$ due to class switching.

\begin{figure}[h]
	\centering
	\includegraphics[width=0.52\textwidth]{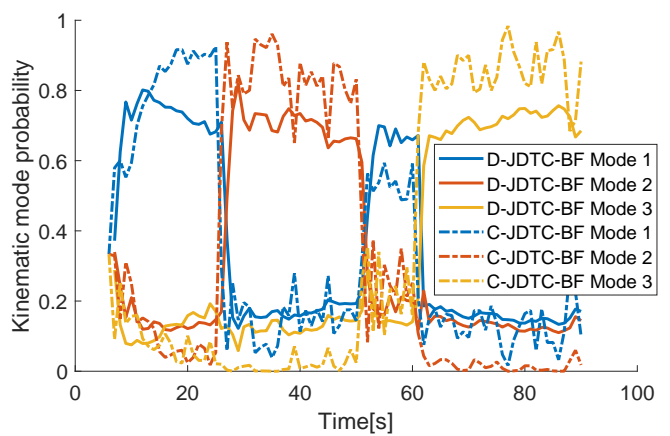}\\
	\caption{Mode probabilities for class 2}
	\label{fig4}
\end{figure}
\begin{figure}[h]
	\centering
	\includegraphics[width=0.52\textwidth]{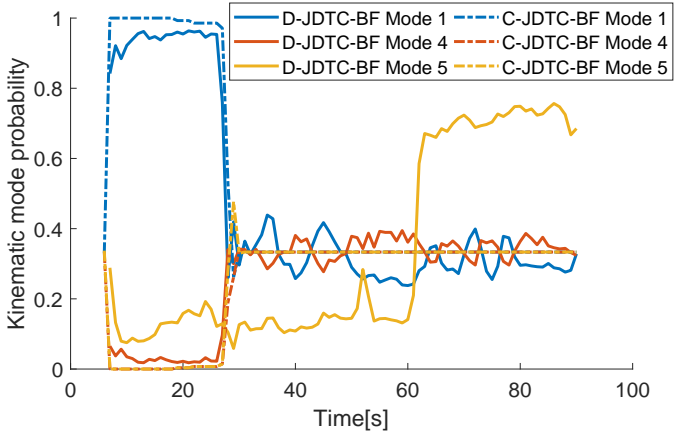}\\
	\caption{Mode probabilities for class 3}
	\label{fig5}
\end{figure}

Estimated mode PMFs for classes 2 and 3 versus time are plotted in Figs. \ref{fig4} and \ref{fig5} (recall that class 1 has just a single mode).
As expected,
both C-JDTC-BF and D-JDTC-BF switch mode whenever the target turning rate changes.
Since all three classes include mode 1,
during the time interval $6-25[s]$ mode 1 is estimated;
then a switching to mode 2 occurs at time $t=26[s]$ (see Fig. {\ref{fig4}}).

\begin{figure}[h]
	\centering
	\includegraphics[width=0.52\textwidth]{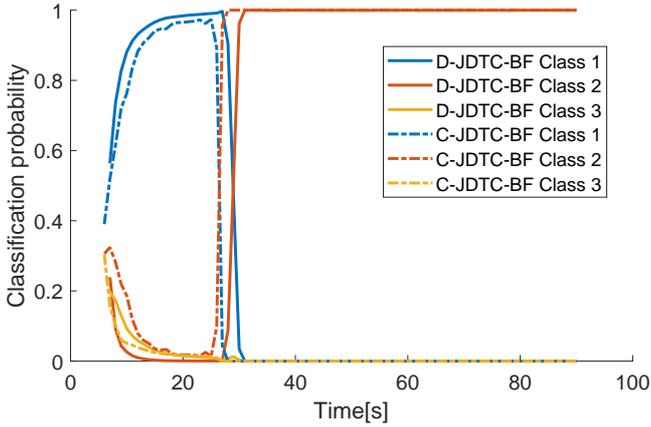}\\
	\caption{Classification results}
	\label{fig6}
\end{figure}
Classification results are shown in Fig. \ref{fig6}.
First, as expected, the estimated class quickly converges to $c_1$
when the target appears in the monitored area. 
This is due to the transition probability of mode 1 in class 1 being greater than that of mode 1 in class 2.
At time $t = 26[s]$, a switching
from class 1 to class 2 occurs as soon as the target starts to behave according to mode 2. 
Before class switching, 
D-JDTC-BF exhibits faster class convergence than C-JDTC-BF, 
while the opposite phenomenon is observed after class switching.

\section{Conclusion}
\textit{Joint detection, tracking and classification} (JDTC) of a single-target immersed in clutter has been addressed by multi-sensor fusion.
The problem has been formulated in a Bayesian framework by introduction
of a suitably augmented Bernoulli density describing the joint distribution of target class, mode and state.
Then, the posed augmented \textit{Bernoulli filtering} (BF) problem has been solved, in both centralized (C-JDTC-BF) and distributed (D-JDTC-BF) settings, 
and a Gaussian-mixture implementation of both filters has been presented.
Simulation experiments have demonstrated the effectiveness of the proposed multi-sensor JDTC-BF approach.
Future work will extend multi-sensor JDTC to multiple maneuvering targets by exploiting \textit{labeled multi-Bernoulli} (LMB) filtering.

\appendix
\section{Proof of Proposition 1}

At time $k$,
we assume that the prior density $f_{k-1}(\mathcal{X})$ is given as 
\begin{equation}
f_{k-1} (\mathcal{X}) = \left\{ \begin{array}{ll} 1-r_{k-1}, & \mbox{if } \mathcal{X} = \emptyset \\
r_{k-1} s_{k-1} ( \mathbf{x} ), & \mbox{if } \mathcal{X} = \{ \mathbf{x} \} \\
0, & \mbox{if } |\mathcal{X} | > 1 \, 
\end{array} 
\right. 
\label{APBD}
\end{equation}
According to \cite[Eqs. 49-51] {yang2015joint},
(\ref{eq3}),
(\ref{eq4}) and (\ref{eq7}),
if $\mathcal{X}= \emptyset$,
we get 
\begin{align}
f_{k|k-1}(\emptyset)&=(1-p_B) \left( 1-r_{k-1} \right)  \nonumber\\
&+r_{k-1} \left[ 1- p_S \sum\limits_{c \in \mathcal{C}} \sum\limits_{m \in \mathcal{M}_c} \int  s_{k-1} (x,c,m) dx \right].
\end{align}
Since $f_{k|k-1}(\emptyset)=1-r_{k|k-1}$,
the predicted EP is given by
\begin{align}
&r_{k|k-1} \nonumber \\
&= p_B(1-r_{k-1}) \nonumber\\
&+ p_S \, r_{k-1} \sum\limits_{c \in \mathcal{C}} \sum\limits_{m \in \mathcal{M}_c}\int s_{k-1}(x,c,m) dx \nonumber \\
&=p_B(1-r_{k-1}) \nonumber\\
&+ p_S \, r_{k-1} \sum\limits_{c \in \mathcal{C}}\gamma_{k-1}(c) \sum\limits_{m \in \mathcal{M}_c}\beta_{k-1}(m|c)\int s_{k-1}(x|c,m) dx. \label{eqA3}
\end{align}
In a similar way,
if $\mathcal{X}= \lbrace \mathbf{x} \rbrace$,
the predicted density can be obtained as
\begin{align}
	&f_{k|k-1}(\mathcal{X})=p_B(1-r_{k-1})\sum\limits_{c \in \mathcal{C}} \sum\limits_{m \in \mathcal{M}_c}s_B(x,c,m) \nonumber \\
	&+ p_S r_{k-1}\sum\limits_{c \in \mathcal{C}} \sum\limits_{m \in \mathcal{M}_c}\int  \Phi_{k|k-1}(x,c,m |x',c,m') \nonumber \\
	& \times s_{k-1}(x',c,m')dx'.
\end{align}

Since the predicted augmented SPDF can be written as $s_{k|k-1}(x,c,m)=\gamma_{k|k-1}(c)\beta_{k|k-1}(m|c)s_{k|k-1}(x|c,m)$,
we can get (\ref{eqA5}), from which the predicted class\&mode-conditioned SPDF $s_{k|k-1}(x|c,m)$ is derived in (\ref{eqA6}).
\begin{figure*}[!t]
	\normalsize
	\begin{align}
		&s_{k|k-1}(x,m,c) \nonumber\\
		&=\gamma_{k|k-1}(c)\beta_{k|k-1}(m|c)s_{k|k-1}(x|m,c) \nonumber\\
		&=\frac{1}{r_{k|k-1}} \left( p_Bs_B(x,m,c)(1-r_{k-1})+r_{k-1}\int p_S \Phi_{k|k-1}(x,m,c |x',m',c) s_{k-1}(x',m',c)dx'\right)\nonumber \\
		&=\frac{1}{r_{k|k-1}} p_B(1-r_{k-1})\gamma_B(c)\beta_B(m|c)s_B(x|m,c)\nonumber\\
		&+\frac{1}{r_{k|k-1}}\left(r_{k-1}\gamma_{k-1}(c)\sum\limits_{m' \in \mathcal{M}_c}\pi_c(m|m')\beta_{k-1}(m'|c)\int p_S \varphi(x|x',m) s_{k-1}(x'|m',c)dx' \right) , \label{eqA5}\\
		&s_{k|k-1}(x|m,c) \nonumber\\
		&=\frac{1}{r_{k|k-1}\gamma_{k|k-1}(c)\beta_{k|k-1}(m|c)} p_B(1-r_{k-1})\gamma_B(c)\beta_B(m|c)s_B(x|m,c)\nonumber\\
		&+\frac{1}{r_{k|k-1}\gamma_{k|k-1}(c)\beta_{k|k-1}(m|c)}\left(r_{k-1}\gamma_{k-1}(c)\sum\limits_{m' \in \mathcal{M}_c}\pi_c(m|m')\beta_{k-1}(m'|c)\int p_S \varphi(x|x',m) s_{k-1}(x'|m',c)dx' \right) .\label{eqA6}	
	\end{align} 
	\hrulefill
	\vspace*{4pt}
\end{figure*}
  
The joint class\&mode PMF $\beta_{k|k-1}(m|c) \gamma(c)$  is then derived as
\begin{align}
&\gamma_{k|k-1}(c) \, \beta_{k|k-1}(m|c)=\int s_{k|k-1}(x,m,c)dx \nonumber \\
&=\frac{p_B(1-r_{k-1})}{r_{k|k-1}}\gamma_B(c)\beta_B(m|c)\int s_B(x|m,c)dx \nonumber\\ 
&+\frac{ p_Sr_{k-1}}{r_{k|k-1}}  \gamma_{k-1}(c)\sum\limits_{m' \in \mathcal{M}_c}\pi_c(m|m')\beta_{k-1}(m'|c) \nonumber \\
& \times \iint {\cal G}(x; f(x',m), Q(m)) s_{k-1}(x'|m',c)dx'dx  \nonumber \\
&=\frac{p_B(1-r_{k-1})}{r_{k|k-1}}\gamma_B(c)\beta_B(m|c) \nonumber \\ &+\frac{p_Sr_{k-1}}{r_{k|k-1}} \left(\gamma_{k-1}(c)\sum\limits_{m' \in \mathcal{M}_c}\pi_c(m|m')\beta_{k-1}(m'|c)  \right) \label{eqA7}
\end{align}
by marginalizing the state $x$ out of the predicted PDF $s_{k|k-1}(x,m,c)$.
Next, from  (\ref{eqA7}) $\beta_{k|k-1}(m|c)$ is derived  as
\begin{align}
&\beta_{k|k-1}(m|c)  =\frac{p_B(1-r_{k-1})}{r_{k|k-1}\gamma_{k|k-1}(c)}\gamma_B(c)\beta_B(m|c) \nonumber \\ &+\frac{p_Sr_{k-1}}{r_{k|k-1}\gamma_{k|k-1}(c)} \gamma_{k-1}(c)\sum\limits_{m' \in \mathcal{M}_c}\pi_c(m|m')\beta_{k-1}(m'|c)    \label{eqA8} 
\end{align}
while $\gamma_{k|k-1}(c)$ is derived as 
\begin{align}
&\gamma_{k|k-1}(c)=\sum\limits_{m \in \mathcal{M}_c}\beta_{k|k-1}(m|c)  \nonumber \\
&=\frac{p_B(1-r_{k-1})}{r_{k|k-1}}\gamma_B(c)+\frac{p_Sr_{k-1}}{r_{k|k-1}} \gamma_{k-1}(c)  \label{eqA9}
\end{align} via marginalization with respect to $c$, i.e. summation over ${\cal M}_c$.

\section{Proof of Proposition 2}

At time $k$, starting from
\begin{equation}
f_{k|k-1} (\mathcal{X}) = \left\{ \begin{array}{ll} 1-r_{k|k-1}, & \mbox{if } \mathcal{X} = \emptyset \\
r_{k|k-1} s_{k|k-1} ( \mathbf{x} ), & \mbox{if } \mathcal{X} = \{ \mathbf{x} \} \\
0, & \mbox{if } |\mathcal{X} | > 1
\end{array}
\right.
\label{PBD}
\end{equation}
the updated density of the Bernoulli RFS $\mathcal{X}$ can be obtained by exploiting the Bayes rule 
\begin{align}
f_{k|k}(\mathcal{X})&=\frac{\ell_k(\mathcal{Z}_k|\mathcal{X}) \, f_{k|k-1}(\mathcal{X})}{\int \ell_k(\mathcal{Z}_k|\mathcal{X}') \, f_{k|k-1}(\mathcal{X}') \, \delta \mathcal{X}'}, \label{eqB1}
\end{align}
where ${\cal Z}_k = \cup_{i \in {\cal N}} {\cal Z}_k^i$.
Supposing that the sensor measurements are conditionally independent,
the multi-sensor likelihood function is factored as
\begin{align}
\ell_k(\mathcal{Z}_k|\mathcal{X})&=\prod\limits_{i \in \mathcal{N}}\ell_k(\mathcal{Z}^{i}_k| \mathcal{X}) 
\end{align}
where $\ell_k(\mathcal{Z}^{i}_k|\mathcal{X})$ is the local likelihood function of node $i$.

According to \cite[Eqs. 7,52-60]{yang2015joint} and (\ref{MS_EP_Context})-(\ref{MS_SPDF_GM_Context}),
we can obtain the normalization constant as (\ref{eqB3}),
\begin{figure*}[!t]
	\normalsize
	\begin{align}
	f_k(\mathcal{Z}_k,\mathcal{Z}_{1:k-1}) &=\int \ell_k(\mathcal{Z}_k|\mathcal{X}) \, f_{k|k-1}(\mathcal{X}) \, \delta \mathcal{X} \nonumber \\
	&=\ell_k(\mathcal{Z}_k|\emptyset ) \, f_{k|k-1}(\emptyset) +r_{k|k-1} \sum\limits_{c \in \mathcal{C}}\gamma_{k|k-1}(c) \sum\limits_{m \in \mathcal{M}_c}\beta_{k|k-1}(m|c) \int \ell_k(\mathcal{Z}_k| \{ x\}) \, f_{k|k-1}(\{ x \}|\mathcal{Z}_{1:k-1}) \, dx \nonumber \\
	&=K(\mathcal{Z}_k) \left[(1-r_{k|k-1})+r_{k|k-1}\sum\limits_{c \in \mathcal{C}}\gamma_{k|k-1}(c) \,\sum\limits_{m \in \mathcal{M}_c} \beta_{k|k-1}(m|c)  \int s_{k|k-1}(x|c,m)\ell(x|c,m)dx \right ]  \nonumber \\
	&=\prod\limits_{i \in \mathcal{N}}K(\mathcal{Z}^i_k) \left[(1-r_{k|k-1})+r_{k|k-1}\sum\limits_{c \in \mathcal{C}}\gamma_{k|k-1}(c)\sum\limits_{m \in \mathcal{M}_c} \beta_{k|k-1}(m|c) \int s_{k|k-1}(x|c,m)\prod\limits_{i \in \mathcal{N}}\ell^i(x|c,m)dx  \right] \label{eqB3}
	\end{align}
	\hrulefill
	\vspace*{4pt}
\end{figure*}
where 
\begin{align}
 \ell^i(x|c,m) &=1-p^i_D(c) +\sum\limits_{z \in \mathcal{Z}^i_k}\frac{K(\mathcal{Z}^i_k \backslash \lbrace{z \rbrace})}{K(\mathcal{Z}^i_k)} \cdot \nonumber \\
 &\cdot  p^i_D(c)\ell^i(z|x),
  \label{eqB5}
\end{align}
and $K(\cdot)$ is the clutter Poisson PDF defined as
$$
K(\mathcal{Z}^i_k)=
e^{\lambda}\prod\limits_{z \in \mathcal{Z}_k^i} \kappa(z)
$$
where $\lambda$ is the expected number of clutter (false) measurements.
Thus, it is immediate to see that
the ratio $K(\mathcal{Z}^i_k \backslash \lbrace z \rbrace)/K(\mathcal{Z}^i_k)$ can be simplified as $1/\kappa(z)$.
Hence,
(\ref{eqB5}) can be rewritten as
\begin{align}
\ell^i(x|c,m)&=1-p^i_D(c) +p^i_D(c)\sum\limits_{z \in \mathcal{Z}^i_k}\frac{\ell^i(z|x)}{\kappa(z)}
\end{align}
where $\ell^i(z|x)$ is given by (\ref{lik}).

By substituting \cite[Eqs. 7]{yang2015joint},(\ref{eqA3}),
(\ref{eqA9}),
(\ref{eqA8}) and (\ref{eqA6}) into (\ref{eqB1}) and taking $\mathcal{X}=\emptyset$,
we get
\begin{align}
	&f_k(\emptyset)=\frac{\ell_k(\mathcal{Z}|\emptyset) \, f_{k|k-1}(\emptyset)}{f_k(\mathcal{Z}_k,\mathcal{Z}_{1:k-1})} \nonumber \\
	&=\frac{\prod\limits_{i \in \mathcal{N}}K(\mathcal{Z}^i_k)(1-r_{k|k-1})}{\prod\limits_{i \in \mathcal{N}}K(\mathcal{Z}^i_k) \left[ (1-r_{k|k-1})+r_{k|k-1}\sum\limits_{c \in \mathcal{C}}\gamma_{k|k-1}(c) \ell(c)\right]  }  \nonumber \\
	&=\frac{(1-r_{k|k-1})}{  (1-r_{k|k-1})+r_{k|k-1}\sum\limits_{c \in \mathcal{C}}\gamma_{k|k-1}(c) \ell(c) }
\end{align}
where 
\begin{align}
\ell(c)&=\sum\limits_{m \in \mathcal{M}_c} \beta_{k|k-1}(m|c) \, \ell(m|c) \label{cl}\\
\ell(m|c)&=\int s_{k|k-1}(x|c,m)\ell(x|c,m)dx \\
\ell(x|c,m)&=\prod\limits_{i \in \mathcal{N}}\ell^i(x|c,m).
\end{align}
Since $f_k(\emptyset)=1-r_k$,
the updated EP is given by
\begin{align}
	r_k=\frac{r_{k|k-1}\sum\limits_{c \in \mathcal{C}}\gamma_{k|k-1}(c) \ell(c)}{  (1-r_{k|k-1})+r_{k|k-1}\sum\limits_{c \in \mathcal{C}}\gamma_{k|k-1}(c) \ell(c)}.
\end{align}
Conversely, when $\mathcal{X}=\lbrace \mathbf{x} \rbrace$
we have
\begin{align}
	&f_k(\lbrace \mathbf{x} \rbrace)=\frac{\ell_k(\mathcal{Z}_k |\lbrace \mathbf{x} \rbrace ) \, f_{k|k-1}(\lbrace \mathbf{x} \rbrace)}{f_k(\mathcal{Z}_k,\mathcal{Z}_{1:k-1})} \nonumber \\
	&=\prod\limits_{i \in \mathcal{N}}K(\mathcal{Z}^i_k)\prod\limits_{i \in \mathcal{N}} \left( 1-p^i_D(c) +p^i_D(c)\sum\limits_{z \in \mathcal{Z}^i_k}\frac{\ell^i(z|x)}{\kappa(z)}\right) \nonumber \\ 
	& \times \frac{r_{k|k-1}\gamma_{k|k-1}(c)\beta_{k|k-1}(m|c)s_{k|k-1}(x|c,m) }{ \prod\limits_{i \in \mathcal{N}}K(\mathcal{Z}^i_k)\left[ (1-r_{k|k-1})+r_{k|k-1}\sum\limits_{c \in \mathcal{C}}\gamma_{k|k-1}(c) \ell(c) \right]} \nonumber \\
	&=\frac{r_{k|k-1}\gamma_{k|k-1}(c)\beta_{k|k-1}(m|c)s_{k|k-1}(x|c,m) \prod\limits_{i \in \mathcal{N}}\ell^i(x|c,m)}{(1-r_{k|k-1})+r_{k|k-1}\sum\limits_{c \in \mathcal{C}}\gamma_{k|k-1}(c) \ell(c)}. \label{eqB9}
\end{align}
Considering that
$$f_k(\lbrace \mathbf{x} \rbrace)\!\!=\!\! \frac{r_k\gamma_{k|k-1}(c)\beta_{k|k-1}(m|c)s_{k|k-1}(x|c,m)\!\!\!\prod\limits_{i \in \mathcal{N}}\!\!\!\ell^i(x|c,m)}{r_{k|k-1}\sum\limits_{c \in \mathcal{C}}\gamma_{k|k-1}(c) \, \ell(c)}$$
we can obtain
\begin{align}
&\gamma_k(c) \, \beta_k(m|c) \, s_k(x|c,m) \nonumber \\
&=\frac{\gamma_{k|k-1}(c)\beta_{k|k-1}(m|c)s_{k|k-1}(x|c,m)\prod\limits_{i \in \mathcal{N}}\ell^i(x|c,m)}{r_{k|k-1}\sum\limits_{c \in \mathcal{C}}\gamma_{k|k-1}(c) \ell(c)} . \label{A_spdf}
\end{align}
Then, via integration of (\ref{A_spdf}) with respect to $x$,
the joint class$\&$mode PMF is given by
\begin{align}
&\gamma_k(c)\beta_k(m|c) \nonumber \\
&=\frac{\int \gamma_{k|k-1}(c)\beta_{k|k-1}(m|c)s_{k|k-1}(x|c,m)\prod\limits_{i \in \mathcal{N}}\ell^i(x|c,m) dx }{r_{k|k-1} \sum\limits_{c \in \mathcal{C}}\gamma_{k|k-1}(c)\ell(c)}. \label{J_MC}
\end{align}
Next, by substituting (\ref{J_MC}) in (\ref{A_spdf}),
the updated 
SPDFs turn out to be given by 
\begin{align}
&s_k(x|c,m)=\frac{s_{k|k-1}(x|c,m)\prod\limits_{i \in \mathcal{N}}\ell^i(x|c,m) }{\int s_{k|k-1}(x'|c,m)\prod\limits_{i \in \mathcal{N}}\ell^i(x'|c,m) dx} \nonumber \\
&=\frac{s_{k|k-1}(x|c,m)\ell(x|c,m) }{\int s_{k|k-1}(x'|c,m)\ell(x'|c,m) \, dx'} \, .
\end{align}

Moreover,
the CPMF can be calculated by
\begin{align}
\gamma_k(c) &=\frac{1}{r_{k|k-1} \sum\limits_{c \in \mathcal{C}}\gamma_{k|k-1}(c)\ell(c)}  \nonumber \\
& \times \sum\limits_{m \in \mathcal{M}_c}\int \gamma_{k|k-1}(c)\beta_{k|k-1}(m|c)s_{k|k-1}(x|c,m) \cdot \nonumber \\
&\cdot \prod\limits_{i \in \mathcal{N}}\ell^i(x|c,m) dx, \label{CPMF_UP}
\end{align}
and from (\ref{J_MC}) the MPMFs
are given by
\begin{align}
&\beta_k(m|c) \nonumber \\
&=\frac{\int \beta_{k|k-1}(m|c)s_{k|k-1}(x|c,m)\prod\limits_{i \in \mathcal{N}}\ell^i(x|c,m) dx}{\sum\limits_{m \in \mathcal{M}_c}\int \beta_{k|k-1}(m|c)s_{k|k-1}(x|c,m)\prod\limits_{i \in \mathcal{N}}\ell^i(x|c,m) dx }\nonumber \\
&=\frac{\beta_{k|k-1}(m|c)\ell(m|c)}{\sum\limits_{m \in \mathcal{M}_c}\beta_{k|k-1}(m|c)\ell(m|c)}.
\end{align}
Finally, according to (\ref{CPMF_UP}) and (\ref{cl}),
the CPMF can be rewritten as 
\begin{align}
	\gamma_k(c)&=\frac{\gamma_{k|k-1}(c)\ell(c)}{\sum\limits_{c \in \mathcal{C}}\gamma_{k|k-1}(c)\ell(c)} .
\end{align}

\section{Proof of Proposition 3} 

By substituting the augmented SPDF $s(x,c,m)= \gamma(c)\beta(m|c) s(x|c,m) $ into (\ref{GCI_rule}), 
we can obtain (\ref{fusion_rule})
	\begin{figure*}[!t]
		\normalsize
		\begin{align}
			\bar{s}(x,c,m)&=\frac{ \prod\limits_{i \in \mathcal{N}} \left [ \gamma^i(c) \beta^i(m|c) s^i(x|c,m) \right ]^{\omega^i} }{\sum\limits_{c \in {\cal C}} \sum\limits_{m \in \mathcal{M}_c} \int \prod\limits_{i \in \mathcal{N}} \left [ \gamma^i(c) \beta^i(m|c) s^i(x|c,m) \right ]^{\omega^i}dx}  =\frac{ \prod\limits_{i \in \mathcal{N}} [ \gamma^i(c)]^{\omega^i}[ \beta^i(m|c)]^{\omega^i}[ s^i(x|c,m) ]^{\omega^i} }{\sum\limits_{c \in {\cal C}} \sum\limits_{m \in \mathcal{M}_c} \int \prod\limits_{i \in \mathcal{N}} [ \gamma^i(c)]^{\omega^i}[ \beta^i(m|c)]^{\omega^i}[ s^i(x|c,m) ]^{\omega^i}dx} \nonumber \\
			&=\frac{ \prod\limits_{i \in \mathcal{N}} [ \gamma^i(c)]^{\omega^i}\prod\limits_{i \in \mathcal{N}}[ \beta^i(m|c)]^{\omega^i}\prod\limits_{i \in \mathcal{N}}[ s^i(x|c,m) ]^{\omega^i} }{\sum\limits_{c \in {\cal C}} \sum\limits_{m \in \mathcal{M}_c} \prod\limits_{i \in \mathcal{N}}  [ \gamma^i(c)]^{\omega^i}\prod\limits_{i \in \mathcal{N}}[ \beta^i(m|c)]^{\omega^i}\int \prod\limits_{i \in \mathcal{N}}[  s^i(x|c,m)  ]^{\omega^i} d x} \nonumber \\
			&=\frac{\prod\limits_{i \in \mathcal{N}}[ s^i(x|c,m) ]^{\omega^i}}{\int \prod\limits_{i \in \mathcal{N}}[  s^i(x|c,m)  ]^{\omega^i} dx} \nonumber \times 
			\frac{\prod\limits_{i \in \mathcal{N}}[ \beta^i(m|c)]^{\omega^i} \int \prod\limits_{i \in \mathcal{N}}[ s^i(x|c,m) ]^{\omega^i} dx}{\sum\limits_{m \in \mathcal{M}_c} \left\lbrace \prod\limits_{i \in \mathcal{N}}[ \beta^i(m|c)]^{\omega^i} \right\rbrace \int \prod\limits_{i \in \mathcal{N}}[  s^i(x|c,m)  ]^{\omega^i} dx} \nonumber \\
			&\times 
			\frac{\prod\limits_{i \in \mathcal{N}} [ \gamma^i(c)]^{\omega^i} \sum\limits_{m \in \mathcal{M}_c} \left\lbrace \prod\limits_{i \in \mathcal{N}}[ \beta^i(m|c)]^{\omega^i} \right\rbrace \int \prod\limits_{i \in \mathcal{N}}[  s^i(x|c,m)  ]^{\omega^i} dx}{\sum\limits_{c \in {\cal C}} \left\lbrace \prod\limits_{i \in \mathcal{N}} [ \gamma^i(c)]^{\omega^i} \right\rbrace \sum\limits_{m \in \mathcal{M}_c} \left\lbrace \prod\limits_{i \in \mathcal{N}}[ \beta^i(m|c)]^{\omega^i} \right\rbrace \int \prod\limits_{i \in \mathcal{N}}[  s^i(x|c,m)  ]^{\omega^i} dx} \nonumber \\
			&=\bar{s}(x|c,m)  \bar{\beta}(m|c) \bar{\gamma}(c), \label{fusion_rule}
		\end{align} 
		\hrulefill
		\vspace*{4pt}
	\end{figure*}
	wherein $\overline{\gamma}(c)$,
	$\overline{\beta}(m|c)$ and  $\overline{s}(x|m,c)$ are fused CPMF,
	MPMFs and SPDFs,
	respectively.
    Moreover,
	the fused EP is
	given by (\ref{NFEP}),
		\begin{figure*}[!t]
		\normalsize
		\begin{align}
			\bar{r}&= \frac{\prod\limits_{i \in \mathcal{N}}(r^i)^{\omega^i}\sum\limits_{c \in {\cal C}} \sum\limits_{m \in \mathcal{M}_c}\int \prod\limits_{i \in \mathcal{N}}[s^{i}(x|c,m) \beta^i(m|c) \gamma^i(c)]^{\omega^i} dx}{\prod\limits_{i \in \mathcal{N}}(1-r^i)^{\omega^i}+\prod\limits_{i \in \mathcal{N}}(r^i)^{\omega^i}\sum\limits_{c \in {\cal C}} \sum\limits_{m \in \mathcal{M}_c} \int \prod\limits_{i \in \mathcal{N}}[s^{i}(x|c,m) \beta^i(m|c) \gamma^i(c)]^{\omega^i} dx} \nonumber \\
			&= \frac{\prod\limits_{i \in \mathcal{N}}(r^i)^{\omega^i}\sum\limits_{c \in {\cal C}} \left\lbrace \prod\limits_{i \in \mathcal{N}} [ \gamma^i(c)]^{\omega^i} \right\rbrace \sum\limits_{m \in \mathcal{M}_c} \left\lbrace \prod\limits_{i \in \mathcal{N}}[ \beta^i(m|c)]^{\omega^i} \right\rbrace \int \prod\limits_{i \in \mathcal{N}}[  s^i(x|c,m)  ]^{\omega^i} dx}{\prod\limits_{i \in \mathcal{N}}(1-r^i)^{\omega^i}+\prod\limits_{i \in \mathcal{N}}(r^i)^{\omega^i}\sum\limits_{c \in {\cal C}} \left\lbrace \prod\limits_{i \in \mathcal{N}} [ \gamma^i(c)]^{\omega^i} \right\rbrace \sum\limits_{m \in \mathcal{M}_c} \left\lbrace \prod\limits_{i \in \mathcal{N}}[ \beta^i(m|c)]^{\omega^i} \right\rbrace \int \prod\limits_{i \in \mathcal{N}}[  s^i(x|c,m)  ]^{\omega^i} dx} \nonumber \\
			&=\frac{\tilde{r}\sum\limits_{c \in {\cal C}}\tilde{\gamma}(c) \sum\limits_{m \in \mathcal{M}_c} \tilde{\beta}(m|c) \int \tilde{s}(x|c,m) dx}{\tilde{\zeta}+\tilde{r}\sum\limits_{c \in {\cal C}}\tilde{\gamma}(c) \sum\limits_{m \in \mathcal{M}_c} \tilde{\beta}(m|c) \int \tilde{s}(x|c,m) dx},	\label{NFEP} 
		\end{align} 
		\hrulefill
		\vspace*{4pt}
	\end{figure*}
	where 
	\begin{align}
		\tilde{\zeta}&=\prod\limits_{i \in \mathcal{N}}(1-r^i)^{\omega^i}, \tilde{r}=\prod\limits_{i \in \mathcal{N}}(r^i)^{\omega^i} \\
		\tilde{\beta}(m|c)&=\prod\limits_{i \in \mathcal{N}}[ \beta^i(m|c)]^{\omega^i},\tilde{\gamma}(c)=\prod\limits_{i \in \mathcal{N}} [ \gamma^i(c)]^{\omega^i} \\ \tilde{s}(x|c,m)&=\prod\limits_{i \in \mathcal{N}}[  s^i(x|c,m)  ]^{\omega^i} dx.
	\end{align}
	Clearly, the fused augmented SPDF $\overline{s}(x,c,m)$ can be factored as  the product of $\overline{\gamma}(c)$,
	$\overline{\beta}(m|c)$
	and $\overline{s}(x|c,m)$.



\begin{thebibliography}{10}
	\expandafter\ifx\csname url\endcsname\relax
	\def\url#1{\texttt{#1}}\fi
	\expandafter\ifx\csname urlprefix\endcsname\relax\def\urlprefix{URL }\fi
	\expandafter\ifx\csname href\endcsname\relax
	\def\href#1#2{#2} \def\path#1{#1}\fi
	
	\bibitem{khan1994target}
	R.~Khan, B.~Gamberg, D.~Power, J.~Walsh, B.~Dawe, W.~Pearson, D.~Millan, Target
	detection and tracking with a high frequency ground wave radar, IEEE Journal
	of Oceanic Engineering 19~(4) (1994) 540--548.
	
	\bibitem{yocom2011bayesian}
	B.~A. Yocom, B.~R. La~Cour, T.~W. Yudichak, A Bayesian approach to passive
	sonar detection and tracking in the presence of interferers, IEEE Journal of
	Oceanic Engineering 36~(3) (2011) 386--405.
	
	\bibitem{petrovskaya2009model}
	A.~Petrovskaya, S.~Thrun, Model based vehicle detection and tracking for
	autonomous urban driving, Autonomous Robots 26~(2) (2009) 123--139.
	
	\bibitem{robin2016multi}
	C.~Robin, S.~Lacroix, Multi-robot target detection and tracking: taxonomy and
	survey, Autonomous Robots 40~(4) (2016) 729--760.
	
	\bibitem{clarke1996detection}
	J.~C. Clarke, A.~Zisserman, Detection and tracking of independent motion, Image
	and Vision Computing 14~(8) (1996) 565--572.
	
	\bibitem{vo2012multi}
	B.~T. Vo, C.~M. See, N.~Ma, W.~T. Ng, Multi-sensor joint detection and tracking
	with the Bernoulli filter, IEEE Transactions on Aerospace and Electronic
	Systems 48~(2) (2012) 1385--1402.
	
	\bibitem{bar2005tracking}
	Y.~Bar-Shalom, T.~Kirubarajan, C.~Gokberk, Tracking with classification-aided
	multiframe data association, IEEE Transactions on Aerospace and Electronic
	Systems 41~(3) (2005) 868--878.
	
	\bibitem{gordon2002efficient}
	N.~J. Gordon, S.~Maskell, T.~Kirubarajan, Efficient particle filters for joint
	tracking and classification, in: Signal and Data Processing of Small Targets
	2002, Vol. 4728, International Society for Optics and Photonics, 2002, pp.
	439--449.
	
	\bibitem{challa2001joint}
	S.~Challa, G.~W. Pulford, Joint target tracking and classification using radar
	and ESM sensors, IEEE Transactions on Aerospace and Electronic Systems 37~(3)
	(2001) 1039--1055.
	
	\bibitem{yang2014joint}
	W.~Yang, Y.~Fu, X.~Li, Joint target tracking and classification via RFS-based
	multiple model filtering, Information Fusion 18 (2014) 101--106.
	
	\bibitem{yang2015joint}
	W.~Yang, Z.~Wang, Y.~Fu, X.~Pan, X.~Li, Joint detection, tracking and
	classification of a manoeuvring target in the finite set statistics
	framework, IET Signal Processing 9~(1) (2015) 10--20.
	
	
	\bibitem{li2012joint}
	W.~ Yang, Y.-W. Fu, X. Li, J.-Q. Long, Joint detection, tracking and
	classification algorithm for multiple maneuvering targets based on
	LGJMS-GMPHDF, Journal of Electronics \& Information Technology 34 (2) (2012) 398-403.
    
	
	\bibitem{yang2013joint}
	W.~Yang, Y.-W. Fu, X.~Li,  Joint detection, tracking and classification of
	multiple maneuvering targets based on the linear Gaussian jump Markov
	probability hypothesis density filter, Optical Engineering 52~(8) (2013)
	083106.
	
	\bibitem{li2016multi}
	M.~Li, Z.~Jing, P.~Dong, H.~Pan, Multi-target joint detection, tracking and
	classification using generalized labeled multi-Bernoulli filter with Bayes
	risk, in: 2016 19th International Conference on Information Fusion,
	IEEE, 2016, pp. 680--687.
	
	\bibitem{gao2016extensions}
	L.~Gao, W.~Sun, P.~Wei, Extensions of the CeMeMBer filter for joint detection,
	tracking, and classification of multiple maneuvering targets, Digital Signal
	Processing 56 (2016) 35--42.
	
	\bibitem{wei2012joint}
	Y.~Wei, F.~Yaowen, L.~Jianqian, L.~Xiang, Joint detection, tracking, and
	classification of multiple targets in clutter using the PHD filter, IEEE
	Transactions on Aerospace and Electronic Systems 48~(4) (2012) 3594--3609.
	
	\bibitem{ilic2013adaptive}
	N.~Ili{\'c}, M.~S. Stankovi{\'c}, S.~S. Stankovi{\'c}, Adaptive consensus-based
	distributed target tracking in sensor networks with limited sensing range,
	IEEE Transactions on Control Systems Technology 22~(2) (2013) 778--785.
	
	\bibitem{olfati2008distributed}
	R.~Olfati-Saber, N.~F. Sandell, Distributed tracking in sensor networks with
	limited sensing range, in: 2008 American Control Conference, IEEE, 2008, pp.
	3157--3162.
	
	\bibitem{ristic2013target}
	B.~Ristic, A.~Farina, Target tracking via multi-static Doppler shifts, IET
	Radar, Sonar \& Navigation 7~(5) (2013) 508--516.
	
	\bibitem{guldogan2014consensus}
	M.~B. Guldogan, Consensus Bernoulli filter for distributed detection and
	tracking using multi-static Doppler shifts, IEEE Signal Processing Letters
	21~(6) (2014) 672--676.
	
	\bibitem{uney2013distributed}
	M.~{\"U}ney, D.~E. Clark, S.~J. Julier, Distributed fusion of PHD filters via
	exponential mixture densities, IEEE Journal of Selected Topics in Signal
	Processing 7~(3) (2013) 521--531.
	
	\bibitem{battistelli2015distributed}
	G.~Battistelli, L.~Chisci, C.~Fantacci, A.~Farina, R.~P. Mahler, Distributed
	fusion of multitarget densities and consensus PHD/CPHD filters, in: Signal
	processing, sensor/information fusion, and target recognition XXIV, Vol.
	9474, International Society for Optics and Photonics, 2015, p. 94740E.
	
	\bibitem{wang2016distributed}
	B.~Wang, W.~Yi, R.~Hoseinnezhad, S.~Li, L.~Kong, X.~Yang, Distributed fusion
	with multi-Bernoulli filter based on generalized covariance intersection,
	IEEE Transactions on Signal Processing 65~(1) (2016) 242--255.
	
	\bibitem{clark2010robust}
	D.~Clark, S.~Julier, R.~Mahler, B.~Ristic, Robust multi-object sensor fusion
	with unknown correlations, IET Sensor Signal Processing for Defence, 2010 p. 14.
	
	\bibitem{mahler2007statistical}
	R.~P. Mahler, Statistical multisource-multitarget information fusion, Artech
	House, Inc., 2007.
	
	\bibitem{vo2005sequential}
	B.-N. Vo, S.~Singh, A.~Doucet, Sequential Monte Carlo methods for multitarget
	filtering with random finite sets, IEEE Transactions on Aerospace and
	Electronic Systems 41~(4) (2005) 1224--1245.
	
	\bibitem{vo2007analytic}
	B.-T. Vo, B.-N. Vo, A.~Cantoni, Analytic implementations of the cardinalized
	probability hypothesis density filter, IEEE Transactions on Signal Processing
	55~(7) (2007) 3553--3567.
	
	\bibitem{mahler2000optimal}
	R.~P. Mahler, Optimal/robust distributed data fusion: a unified approach, in:
	Signal Processing, Sensor Fusion, and Target Recognition IX, Vol. 4052,
	International Society for Optics and Photonics, 2000, pp. 128--138.
	
	\bibitem{xiao2005scheme}
	L.~Xiao, S.~Boyd, S.~Lall, A scheme for robust distributed sensor fusion based
	on average consensus, in: IPSN 2005. Fourth International Symposium on
	Information Processing in Sensor Networks, 2005., IEEE, 2005, pp. 63--70.
	
	\bibitem{battistelli2013consensus}
	G.~Battistelli, L.~Chisci, C.~Fantacci, A.~Farina, and A.~Graziano, Consensus
	CPHD filter for distributed multitarget tracking, IEEE Journal of
		Selected Topics in Signal Processing, 7~(3) (2013) 508--520.
		
       \bibitem{orguner}
       M. G\"unay, U. Orguner, and M. Demirekler, 
       Chernoff fusion of Gaussian mixtures. based on sigma-point approximation, IEEE Transactions on Aerospace and Electronic Systems, 
       52~(6) (2016) 2032-2046.       
	
	\bibitem{battistelli2015consensus}
	G.~Battistelli, L.~Chisci, C.~Fantacci, A.~Farina, A.~Graziano, Consensus-based
	multiple-model Bayesian filtering for distributed tracking, IET Radar, Sonar
	\& Navigation 9~(4) (2015) 401--410.
	
	\bibitem{vo2006gaussian}
	B.-N. Vo and W.-K. Ma, ``The Gaussian mixture probability hypothesis density
	filter,'' IEEE Transactions on Signal Processing, vol.~54, no.~11, pp.
	4091--4104, 2006.
	
	\bibitem{julier2006empirical}
	S.~J. Julier, An empirical study into the use of Chernoff information for
	robust, distributed fusion of Gaussian mixture models, in: 2006 9th
	International Conference on Information Fusion.\hskip 1em plus 0.5em minus
	0.4em\relax IEEE, 2006, pp. 1--8.
	
	\bibitem{schuhmacher2008consistent}
	D.~Schuhmacher, B.-T. Vo, B.-N. Vo, A consistent metric for performance
	evaluation of multi-object filters, IEEE Transactions on Signal Processing
	56~(8) (2008) 3447--3457.
	
\end{thebibliography}
\end{document}